\documentclass{article}

\usepackage{microtype}
\usepackage{graphicx}
\usepackage{subcaption}
\usepackage{booktabs} 
\usepackage{tcolorbox}

\usepackage[accepted]{icml2026}

\usepackage{amsmath}
\usepackage{amssymb}
\usepackage{mathtools}
\usepackage{amsthm}
\usepackage{algorithmicx}
\usepackage[noEnd=true, indLines=true]{algpseudocodex}
\usepackage{threeparttable}

\usepackage{multirow}
\usepackage{makecell}
\usepackage{caption} 
\tcbuselibrary{breakable,skins}
\usepackage{fvextra} 
\usepackage{pifont}
\tcbset{
  algobox/.style={
    enhanced,
    colback=black!2,        
    colframe=black!60,      
    boxrule=0.8pt,          
    arc=0mm,                
    left=2mm, right=2mm,    
    top=1.5mm, bottom=1.5mm,
    boxsep=0mm,             
    sharp corners=southwest,
    equal height group=algotwins,
  }
}

\newcolumntype{Y}{>{\centering\arraybackslash}X}

\usepackage[capitalize,noabbrev]{cleveref}

\theoremstyle{plain}

\theoremstyle{definition}

\theoremstyle{remark}

\usepackage[textsize=tiny]{todonotes}

\icmltitlerunning{Lessons from Hybrid Multi-Agent Systems}

\begin{document}

\twocolumn[
 \icmltitle{When Cloud Agents Meet Device Agents:\\Lessons from Hybrid Multi-Agent Systems} 

\icmlsetsymbol{equal}{*}

\begin{icmlauthorlist}
  \icmlauthor{Corrado Rainone*}{qualcomm}
  \icmlauthor{Davide Belli*}{qualcomm}
  \icmlauthor{Bence Major}{qualcomm}
  \icmlauthor{Arash Behboodi}{qualcomm}
\end{icmlauthorlist}

   \icmlaffiliation{qualcomm}{Qualcomm AI Research\textsuperscript{\dag}}


\icmlcorrespondingauthor{Corrado Rainone}{crainone@qti.qualcomm.com}

\vskip 0.3in
]

\printAffiliationsAndNotice{\icmlEqualContribution \textsuperscript{\dag}Qualcomm AI Research is an initiative of Qualcomm Technologies, Inc.}

\begin{abstract}
The design space of agentic AI inference spans two extremes: frontier large language models (LLMs), typically hosted in the cloud and offering strong performance across a wide range of tasks at substantially high cost, and more cost-efficient small language models (SLMs), which are amenable to on-device inference.
Hybrid multi-agent systems (MASs) combining on-device and cloud models offer a promising middle ground, but they also introduce a complex and poorly understood design space in which task accuracy, monetary cost, and edge energy consumption are tightly coupled; in the absence of general design principles, hybrid components, although not the most prevalent choice, are typically introduced through ad hoc decisions tailored to specific domains.
In this work, we examine this design space more systematically. We adapt two representative MAS architectures to support hybrid inference and study how individual design choices shift the operating point along the Pareto frontier of power, cost, and performance.
Our findings paint a nuanced picture of hybrid MAS design: while SLMs can effectively benefit from LLM assistance, the optimal architecture is highly task-dependent, and greater frontier-level compute does not consistently translate to better performance.
\end{abstract}

\section{Introduction}
Large Language Models (LLMs) were initially designed and evaluated primarily for short-horizon, turn-based interaction with human users, such as question answering and conversational assistance. Over the past year, however, LLMs have increasingly been deployed as agents: systems that must interpret high-level, underspecified user goals, autonomously synthesize and execute multi-step plans, and operate over extended time horizons while invoking tools, maintaining state, and interacting with an external environment.
Such complex workflows are often challenging even for the most capable frontier models~\cite{Sinha2025-om}. Owing to their scale and corresponding computational demands, these models are typically deployed in centralized cloud infrastructures and accessed through online APIs. These APIs are commonly offered on a subscription model with per-token costs. As a result, users who rely on frontier models to power agentic systems (e.g., OpenClaw~\footnote{\url{https://github.com/openclaw/openclaw}}) may face significant operational costs or rapidly depleted token budgets, particularly due to the inherently long-horizon and iterative nature of agentic workloads.

Language Models can, however, also come in sizes small enough to fit on a consumer laptop, or even a smartphone. Users are increasingly looking into self-hosted or on-device solutions as a way to avoid ballooning API costs, to hedge against the eventuality that frontier models be retired or become suddenly unavailable, or out of concerns for data ``sovereignty'' and privacy. Over the past years, improvements in the quality of training algorithms and advances in test-time compute have led to remarkable improvements in the quality of Small Language Models (SLMs)~\cite{lu2024small}; device-sized models can also be trained or fine-tuned to the point of matching or exceeding the performance of frontier LLMs in restricted domains~\cite{hsieh2023distilling,zhao2024lora,liu2024ddk}, including the agentic one~\cite{Kang2025-lb}. Nevertheless, outside of these restricted domains, a significant gap with frontier LLMs remains. And an even bigger gap, which is particularly relevant to long-horizon performance, can be found when it comes to the \emph{context length} that these models can support: the typical DRAM sizes on consumer devices impose a commensurably low upper bound to KV-cache and context length size, a bound which effectively vanishes when the LM is hosted on the cloud.

A straightforward solution to this problem can come in the form of \emph{Hybrid} systems, e.g. a system in which a router module can choose from a suite of models with different levels of proficiency~\cite{Ding2024-HybridLLM,Stripelis2024-TensorOpera,Ong2024-RouteLLM, zhang2024llmcascade,jitkrittum2026uniroute}, or one in which a task can be escalated to a more proficient model when a less capable one struggles to solve it~\cite{MAIUI}. In this work we seek to explore whether this hybrid philosophy can instead be realized with models that not only have different levels of proficiency, but also take up different \emph{roles} in the solution process.
We implement and study a class of Hybrid \emph{Multi-Agent} Systems (MASs) in which ideas and principles from the MAS literature are adapted to a Hybrid setup spanning both cloud and edge. We explore the space of design choices such as model assignment, agent roles, and amount of cloud compute used, from a standpoint of both performance and efficiency, over tasks involving Deep Search and UI assistance. We make some unexpected observations, and carry out in-depth studies to investigate the mechanisms that are responsible for them. Our main contributions and findings are as follows:
\begin{enumerate}
\item \textbf{Systematic study of hybrid MAS design:} we adapt representative multi-agent architectures to the cloud–edge setting and introduce a unified framework to evaluate trade-offs between accuracy, monetary cost, and edge energy.
\item \textbf{Task-dependent behavior of hybrid architectures:} we show that no single design dominates; plan-based and advisory paradigms excel in different domains, and increasing cloud usage does not consistently improve performance.
\item \textbf{Mechanistic insights into hybrid collaboration:} we identify key factors such as supervision frequency, restart policies, and summarization that explain when hybridization helps or harms long-horizon reasoning.
\item \textbf{Context Efficiency of hybrid MAS:} we show that these systems handle long contexts more effectively through context resets and summarization, which limit KV-cache growth and improve suitability for memory-constrained edge deployment.
\end{enumerate}

\section{Related Work}

\paragraph{Multi-Agent systems}
Recent surveys highlight the rapid proliferation of multi-agent systems built on large language models, spanning diverse design choices in architecture, planning, memory, and communication mechanisms~\cite{wei2024survey,han2024mas,cui2025collaboration,gao2025harmony}. A key axis of variation lies in the \emph{architecture} of a MAS: prior work explores both \emph{centralized} designs \cite{zhang2024chain,wang2024mobile,nayak2024long}, where a planner or orchestrator coordinates specialized agents (e.g., star or tree topologies), and more \emph{decentralized} designs~\cite{yangagentnet,zhu2025multiagentbench,zhang2024cut}, where agents interact through mesh- or graph-like communication patterns. These choices are often intertwined with other dimensions, such as planning paradigms (e.g., ReAct, Chain-of-Thought) and memory mechanisms, resulting in a combinatorial design space rather than a single dominant paradigm.

Building on this diversity, most existing systems are tailored to specific tasks or benchmarks, with architectures, protocols, and evaluation setups co-designed within individual works~\cite{bogavelli2025agentarch,zeng2025e2edev,cemri2024failures}. As a result, relatively few works (e.g.,~\cite{Cemri2025-jl,Kim2025-fq}) systematically evaluate the impact of architectural decisions across a broad range of tasks. Furthermore, recent analyses emphasize that MAS failure modes are highly sensitive to design choices, such as task decomposition, communication protocols, and verification strategies~\cite{zhang2025debate,Cemri2025-jl}, reinforcing their dependence on task-specific factors. Taken together, the literature highlights the breadth of the MAS design space, while providing limited actionable guidance for selecting architectures, thereby motivating the need for more systematic studies of architectural trade-offs.

\begin{figure*}[t]
  \centering

  \includegraphics[width=\textwidth]{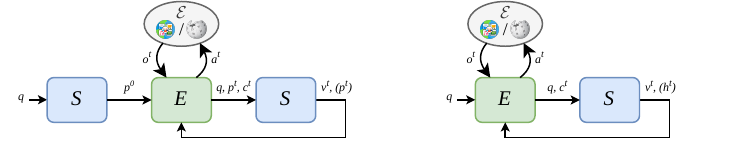}
  \begin{minipage}[t]{0.49\linewidth}
    \centering
    \phantomsection 

    \begin{tcolorbox}[algobox]
      \small
      \begin{algorithmic}[1]
        \Require Query $q$, Supervisor $S$, Executor $E$, \\
                 Environment $\mathcal{E}$, max turns $T$, \\
                 verification interval $T_v$, \\
                 Instructions $I_{\text{plan}}, I_{\text{exec}}, I_{\text{verify-replan}}$
        \State $c^{0} \gets \varnothing$ \Comment{Initialize context (empty)}
        \State $p^{0} \gets S(I_{\text{plan}}, q)$ \Comment{Initial plan}
        \For{$t \gets 1$ to $T$} \Comment{ReAct loop}
            \State $(r^{t}, a^{t}) \gets E(I_{\text{exec}}, q, p^{t-1}, c^{t-1})$
            \State $o^{t} \gets \mathcal{E}(a^{t})$
            \State $c^{t} \gets c^{t-1} + [r^{t}, a^{t}, o^{t}]$
            \State $p^{t} \gets p^{t-1}$
            \If{$t \bmod T_v = 0$} \Comment{Verify}
                \State $v^{t}, (p^{t}_{\text{new}}) \gets S(I_{\text{verify-replan}}, q, p^{t}, c^{t})$
                \If{$v^{t} = 1$} \Comment{Intervene (replan)}
                    \State $p^{t} \gets p^{t}_{\text{new}}$
                \EndIf
            \EndIf
        \EndFor
      \end{algorithmic}
    \end{tcolorbox}
    \vspace{0.2em}
    a) PEVR architecture
  \end{minipage}
  \hfill
  \begin{minipage}[t]{0.49\linewidth}
    \centering
    \phantomsection 
    \begin{tcolorbox}[algobox]
      \small
      \begin{algorithmic}[1]
        \Require Query $q$, Supervisor $S$, Executor $E$, \\
                 Environment $\mathcal{E}$, max turns $T$, \\
                 verification interval $T_v$, \\
                 Instructions $I_{\text{exec}}, I_{\text{verify-advice}}$
        \State $c^{0} \gets \varnothing$ \Comment{Initialize context (empty)}
        \State $h^{0} \gets \varnothing$ \Comment{Initial advice (empty)}
        \For{$t \gets 1$ to $T$} \Comment{ReAct loop}
            \State $(r^{t}, a^{t}) \gets E(I_{\text{exec}}, q, h^{t-1}, c^{t-1})$
            \State $o^{t} \gets \mathcal{E}(a^{t})$
            \State $c^{t} \gets c^{t-1} + [r^{t}, a^{t}, o^{t}]$
            \State $h^{t} \gets h^{t-1}$
            \If{$t \bmod T_v = 0$} \Comment{Verify}
                \State $v^{t}, (h^{t}_{\text{new}}) \gets S(I_{\text{verify-advice}}, q, c^{t})$
                \If{$v^{t} = 1$} \Comment{Intervene (advise)}
                    \State $c^{t} \gets \varnothing$ \Comment{Reset context}
                    \State $h^{t} \gets h^{t}_{\text{new}}$ \Comment{Update advice}
                \EndIf
            \EndIf
        \EndFor
      \end{algorithmic}
    \end{tcolorbox}
    \vspace{0.2em}
    b) EVA architecture
  \end{minipage}

  \vspace{0.4em}

  \caption{System diagrams for the PEVR (Top, a) and EVA (Top, b) Hybrid Multi-Agent Systems, with pseudocode (Bottom) for each architecture. Top: both architectures include a ReAct loop between executor $E$ and environment $\mathcal{E}$, and an outer loop with the supervisor $S$. The PEVR supervisor also generates an initial plan. In case of verifier intervention, the supervisor produces a replan in PEVR and a general advice in EVA. Bottom: in the pseudocode, $r^t$ is the executor reasoning trace, $a^t$ is the action predicted by the executor, $o^t$ is the observation from the environment, $c^t$ is the context with for current ReAct trajectory, $+$ is the string concatenation operator, $v^t \in {0, 1}$ is the verifier outcome (continue or intervene), $p^t_{\text{new}}$ and $h^t_{\text{new}}$ are respectively the replan and the new advice, only generated if the PEVR and EVA verifiers decided to intervene. }
  \label{fig:system-diagram}
\end{figure*}

\paragraph{Hybrid AI} 
An increasingly important line of work, which we refer to as \emph{Hybrid AI} in this paper, investigates how agentic pipelines should allocate computation across models of different sizes and deployment locations. In particular, \citet{belcak2025slmfuture} argue that most agentic calls are sufficiently specialized and repetitive that small language models (SLMs) are often preferable to frontier-scale models. More generally, a growing body of work points to a design space in which small models can be effectively integrated into agentic frameworks. For example, \citet{ToolOrchestra} show that an 8B model can serve as an orchestrator for calling multiple tools, and similarly, \citet{jhandi2025slmtoolcalling} report that a targeted fine-tuned 350M model achieves strong performance on ToolBench-style evaluation. Many recent works have developed frameworks and methods for routing requests across different models: small or large, on-device or in the cloud \cite{Ding2024-HybridLLM,Stripelis2024-TensorOpera,Ong2024-RouteLLM, zhang2024llmcascade,jitkrittum2026uniroute}. 

Deep-search systems are a particularly natural target for hybrid designs, as they must interleave query decomposition, retrieval, browsing, evidence aggregation, and synthesis over long horizons \cite{Tongyi-DeepResearch-Team2025-us,DRVenus-Team2026}. GUI agents provide a similar motivation for hybrid architectures, since they combine perception, grounding, action selection, and multi-step recovery under tight latency and privacy constraints. Recent GUI agents span a wide range of model sizes, from small to large \cite{MAIUI,UI-Venus-1.5,UI-TARS-2}. Notably, MAI-UI introduces a native hybrid-AI design \cite{MAIUI}, in which a cloud model is invoked to recover from failures of the on-device model.

\paragraph{Context summarization and reset}
Context summarization and reset are standard mechanisms in LLM-based multi-agent systems, motivated by both strict context limits and empirically observed degradation effects such as lost-in-the-middle and context rot \cite{Sinha2025-om,liu2024lostmiddle,hong2025contextrot,nadar2025contextdiscipline}. In practice, most systems rely on pre-trained LLMs to summarize interaction histories, tool traces, or intermediate reasoning into compact task state, rather than training dedicated compression models \cite{weng2023llmagents,du2026memoryagents,lu2025summrlcontext}. Recent MAS architectures integrate summarization into replanning in distinct ways: Planner-Executor systems such as AgentFlow incorporate memory updates that retain tool-use trajectories while pruning verbose reasoning \cite{li2025agentflow}, whereas Advisor-style frameworks explicitly reset executor context using corrective summaries generated by stronger planning or critique modules \cite{qin2025advisor}. 
 
\section{Hybrid Multi-Agent architectures}
\label{sec:archs}

In this section, we identify how core principles of Multi-Agent Systems (MAS), such as collaboration, orchestration, role specialization, and tool use, can be adapted to a hybrid cloud-edge inference setting. Our goal is not to introduce new MAS abstractions, but to study how established agentic patterns behave when deployed across models with widely different capabilities, costs, and resource constraints. This leads us to the following design desiderata: \emph{i}) The system must operate across models with substantially different capability profiles, from device-sized models to frontier-scale LLMs - \emph{ii}) Multi-turn execution, which is token intensive, should preferably occur on device to lower cloud API cost - \emph{iii}) Users should have explicit control over cloud involvement, enabling a tunable performance-cost trade-off - \emph{iv}) The context of edge-based agents must remain bounded to respect DRAM and KV-cache constraints.

A common organizing principle in the MAS literature is \emph{orchestration}, where a higher-level agent decomposes a task and supervises one or more Executor agents responsible for sustained interaction with tools and environments. This pattern enforces a clear separation of roles: long-horizon, token-heavy execution is decoupled from lighter-weight planning, verification, and redirection. Such a separation is particularly well suited to hybrid settings, as it allows expensive cloud models to act sparingly and strategically, while lower-cost edge models handle execution. Guided by this observation and the desiderata above, we study two representative hybrid agentic architectures: PEVR and EVA. We provide in Figure~\ref{fig:system-diagram} the system diagrams and pseudocode for both architectures.

\paragraph{Plan–Execute–Verify–Replan (PEVR)}
In PEVR, a cloud-based Supervisor first receives the user query and produces a natural-language execution plan, which is passed to an on-device Executor. During execution, the Supervisor periodically verifies whether the Executor’s actions remain aligned with the original plan. Upon detecting misalignment or failure, it intervenes by issuing a revised plan that specifies the remaining steps or redirects execution. The Executor is then re-prompted to continue from this updated plan. This architecture most strongly reflects centralized orchestration, with the Supervisor defining both the initial strategy and any subsequent course corrections. Its design is closely related to the AgentFlow architecture proposed by \citet{Li2025-ms}, and shows strong similarities to MAS architectures for GUI assistants \cite{wang2024mobile} and robotic navigation \cite{nayak2024long}.

\paragraph{Execute–Verify–Advise (EVA)}
In EVA, the on-device Executor directly receives the user query and proceeds via a ReAct-style loop~\cite{Yao2022-nl}, without an explicit initial plan. A cloud-based Supervisor periodically verifies task progress and intervenes only when execution appears stalled or incorrect. Instead of replanning, the Supervisor provides (i) a summary of the Executor’s prior actions and (ii) advisory feedback describing how to proceed and avoid past errors. Compared to PEVR, supervision is based on observed progress rather than plan adherence, and redirection takes the form of guidance rather than an explicit plan. This design is closely related to advisor-style architectures \cite{claude-advisor,ferraz2024llm,asawa2025train}.

Both architectures satisfy our desiderata. In each case, execution (the most token-intensive component) is confined to the edge, while the cloud model performs intermittent, higher-level supervision. The user controls the interval (in steps) between supervisions, and thus cloud costs, via a single verification parameter. Finally, in both architectures, supervisory intervention triggers a context reset for the Executor: past interaction history is replaced with a fresh prompt derived from either a plan or a summary. This prevents unbounded context growth, enables clean recovery from errors, and keeps KV-cache usage within practical limits for edge deployment.

\section{Experimental setup}
Each architecture presented in Section~\ref{sec:archs} represents a way of assigning agentic roles and structuring inter-agent communication. Within it, there are more design decisions to be taken, such as choosing backbone LMs and the frequency of supervision. We now outline the experimental setup and metrics we use to assess their impact.

\subsection{Benchmarks}
We focus on three benchmarks of increasing difficulty: HotpotQA, FanOutQA, and AppWorld. Together, they span short-horizon multi-hop reasoning, long-horizon information aggregation, and stateful interactive decision making.

\paragraph{HotpotQA}
HotpotQA is a multi-hop question answering benchmark that requires retrieving and combining evidence from multiple Wikipedia articles. From an agentic perspective, tasks are short-horizon and involve a small number of reasoning and retrieval steps. Performance is measured using Exact Match and ROUGE${}_1$ F1, and we report the latter in our results. We report results on the first 500 questions of the validation split, fullwiki version.

\paragraph{FanOutQA}
FanOutQA extends multi-hop QA to fan-out queries that require identifying a dynamic set of relevant entities and aggregating information across many documents. Compared to HotpotQA, it induces a longer reasoning horizon and greater demands on decomposition and intermediate state tracking. We evaluate performance using ROUGE${}_1$ .

\paragraph{AppWorld}
AppWorld is an interactive benchmark in which agents execute multi-step programs within a simulated ecosystem of everyday applications via APIs. Tasks are long-horizon and stateful, often requiring planning, control flow, error recovery, and careful execution to avoid unrecoverable side effects (such as sending emails to the wrong recipient). Performance is evaluated using Test Pass Ratio and Task Success (all tests passing) based on the final environment state.

Additional details on these benchmarks and their respective environments and tools are provided in Appendix~\ref{app:benchmarks}.

\begin{figure*}[t]
  \centering
  \includegraphics[width=\textwidth]{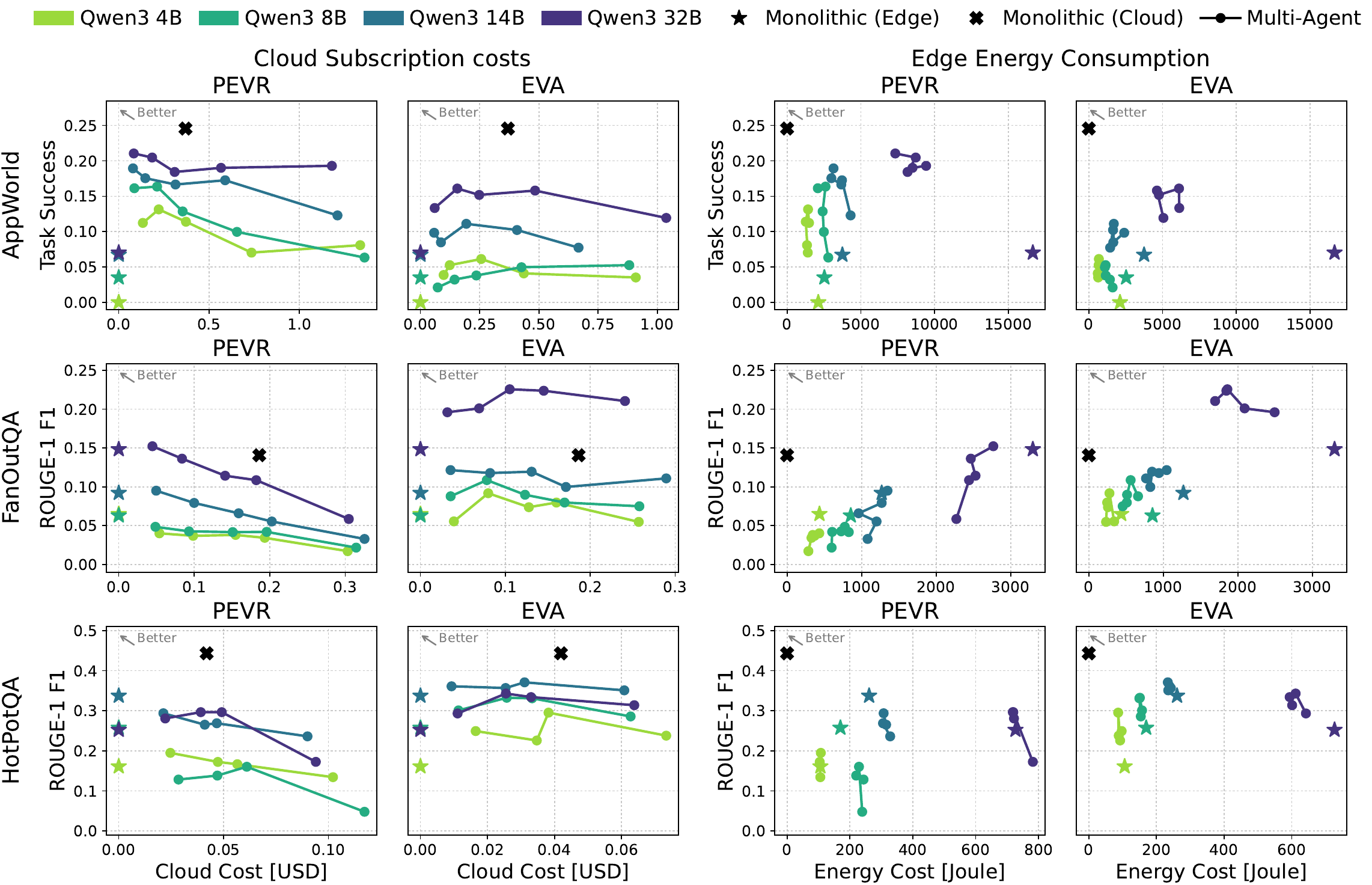}
  \caption{Experimental results comparing Monolithic systems (edge-only and cloud-only) against Multi-Agent systems (PEVR and EVA). We present one row per benchmark, compare performance against both API cost in \$ (on the two leftmost columns), and energy cost in Joules (on the two rightmost columns). For MASs, each line corresponds to a sweep on the verification interval.} 
  \label{fig:main-plot}
\end{figure*}

\subsection{Efficiency metrics}
We focus on two efficiency metrics. API cost in dollars for cloud agents, and energy consumption, measured in Joules, for edge agents. For each cloud LM inference, we can calculate its API cost using the pricing tables of the associated models (see Appendix \ref{app:cloud}). For edge LM inference, we estimate its energy consumption using a simple model which we outline in Appendix~\ref{app:energy-model}; we therefore do not actually deploy our LMs on device and measure their energy consumption. 

In both cases, we obtain the cost of a multi-turn trajectory by adding the costs of the LLM inferences at each turn. We also evaluate the context efficiency by tracking the maximum KV-cache footprint over the task trajectory, calculated as described in Appendix~\ref{app:kv}).

\subsection{MAS backbones and hyper-parameters}
\label{subsec:models}
We choose GPT-4o~\cite{openai_gpt4o_2024} as our on-cloud agent. For edge models, we pick the 4B, 8B, 14B, and 32B variants of the Qwen3 series~\cite{qwen3_technical_report_2025}. We choose these because they are small enough to be deployable on edge devices; The 4B and 8B variants are small enough to run on a smartphone, while the 14B and 32B variants can fit on a consumer laptop. We access GPT-4o via Azure, while we use vLLM~\cite{kwon2023vllm} to run inference on models of the Qwen3 series. We employ fp8 quantization of KV-cache and weights for the 32B variant, which allows us to use only one A100 GPU for all of our runs.

In our main experiments, we set the maximum number of ReAct turns to 10 for HotPotQA, 20 for FanOutQA, and 40 for AppWorld. Based on these values for maximum number of turns, we verify every [1,2,3,5] steps for HotpotQA, [1,2,3,5,10] for FanOutQA, and [1,2,4,8,16] for AppWorld.

\section{Exploring the design space of Hybrid MASs}

We present the main results of this work in Figure~\ref{fig:main-plot}. 
The figure compares the performance of monolithic edge and cloud systems against PEVR and EVA, with varying verification interval from the supervisor model. On the left side, we consider the task performance as a function of the subscription costs for the cloud agent, while on the right side we study the performance against the energy consumption of the edge agent. Each rows presents with four plots the results on each of our three benchmarks.

\begin{figure*}[t]
  \centering
  \includegraphics[width=\textwidth]{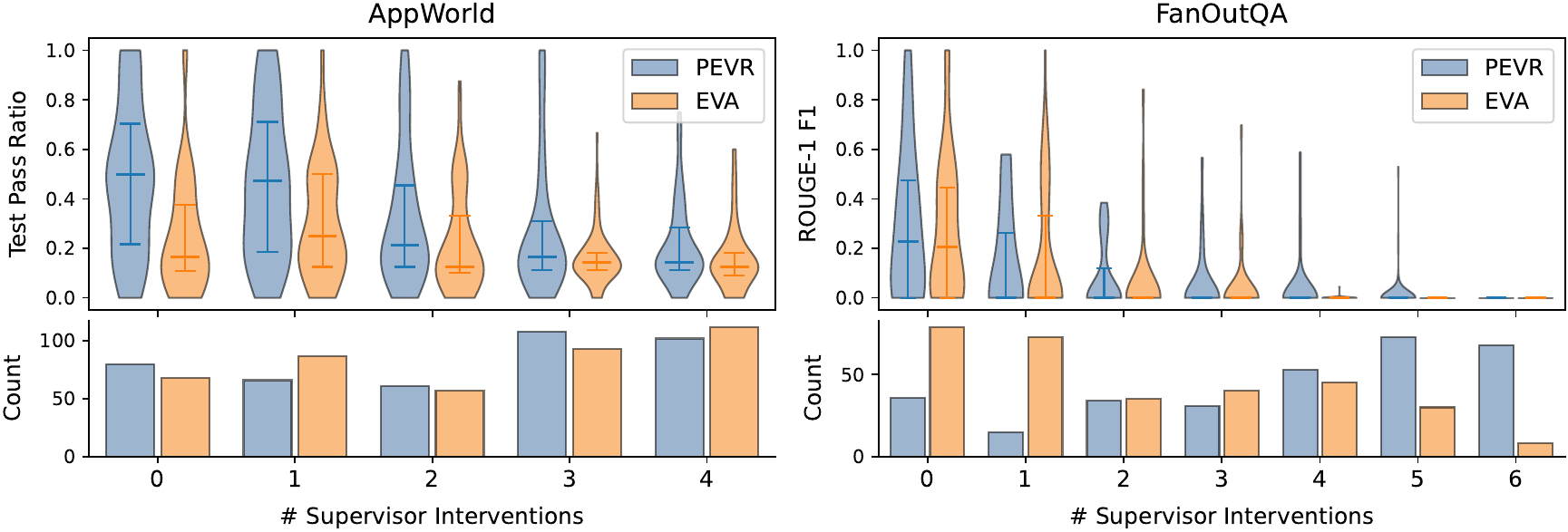}
  \caption{Comparison of PEVR and EVA performance as a function of the number of Supervisor interventions (top), and the distribution of interventions (bottom). Results obtained using Qwen 3 14B as Edge model on AppWorld and FanOutQA, respectively with verification interval of 8 and 3, and max \# of turns 40 and 20. The densities present markers at the quartile values ($25^{th}$, $50^{th}$, $75^{th}$).}
  \label{fig:violin}
\end{figure*}

Some interesting observations can be garnered from Figure~\ref{fig:main-plot}:
\begin{itemize}
\item \textbf{Collaboration between edge and cloud is possible}: Even with device-sized Executor agents, it is always possible to find a MAS architecture and verification interval which delivers better performance than a monolithic on-device agent, and for less cost than a monolithic on-cloud agent.
\item \textbf{There is no ``free lunch'' MAS architecture}: the PEVR variant is clearly superior to the EVA one on the UI assistance domain (AppWorld), but the opposite happens on the Deep Search domain (HotpotQA and FanOutQA).
\item \textbf{Increasing cloud compute usage does to necessarily increase performance}: In both architectures, there often is an optimal verification interval beyond which performance starts to degrade.
\item \textbf{Both PEVR and EVA retain or improve task accuracy} when compared to the Monolithic Edge models, while reducing the energy consumption by up to $3\times$ thanks to the Executor context resets. The EVA variant tends to result in larger energy savings than PEVR, thanks to its summarization feature.
\item \textbf{Executor size correlates positively with performance} in all cases.
\end{itemize}
Overall, we can see that multiple MAS configurations reside on the Pareto frontier of these plots, confirming that Hybrid MASs provide a way to tune the accuracy-cost trade-off for a given task.
In the following of this section, we shed light on the mechanisms behind our main results with a series of in-depth studies.

\begin{table}[t]
  \centering
  \caption{Ablation study on verifier performance based on plan (PEVR) or query (EVA) verification. The plan-based verifier intervenes more aggressively (higher false positives), while the query-based one is more conservative, (higher false negatives). All results are based on Qwen 3 32B. FanOutQA tasks are labeled as successful if ROUGE${}_1$ F1 $> 0.5$.}
  \label{tab:verifier-ablation}
  \resizebox{\linewidth}{!}{%
  \setlength{\tabcolsep}{4pt} 
  \begin{tabular}{ll|cc}
    \toprule
    \makecell{\textbf{Bench.}} & \makecell{\textbf{Verifier}}
    & \makecell{\textbf{False} \\ \textbf{Negative}}
    & \makecell{\textbf{False} \\ \textbf{Positive}} \\
    \midrule

    \multirow{2}{*}{\centering AppWorld}
    & \makecell[l]{Plan-based (PEVR)} & \textbf{5.3\%} & 6.2\% \\
    & \makecell[l]{Query-based (EVA)} & 6.0\% & \textbf{1.9\%} \\
    \midrule

    \multirow{2}{*}{\centering FanOutQA}
    & \makecell[l]{Plan-based (PEVR)} & \textbf{8.4\%} & 7.7\% \\
    & \makecell[l]{Query-based (EVA)} & 14.8\% & \textbf{6.1\%} \\



    \bottomrule
  \end{tabular}
  }
\end{table}

\begin{table*}
\centering 
  \caption{Comparing the previously discussed hybrid setup against a symmetric configuration with cloud Executor and edge Supervisor. This setup is suboptimal both in terms of task accuracy and cloud subscription costs. Scores are obtained with PEVR for AppWorld and EVA for FanOutQA, selecting the verfication interval with best task accuracy. We highlight in bold results residing on the Pareto frontier.}
  \label{tab:reverse}
  \resizebox{0.9\textwidth}{!}{%
  \setlength{\tabcolsep}{8pt} 
    



    \begin{tabular}{ll|cc|cc}
    \toprule
    & & \multicolumn{2}{c|}{\textbf{AppWorld}} & \multicolumn{2}{c}{\textbf{FanOutQA}} \\
    \cmidrule(lr){3-4} \cmidrule(lr){5-6}
    \textbf{Executor} & \textbf{Supervisor}
    & \makecell{\textbf{Task Succ. ($\uparrow$)}}
    & \makecell{\textbf{Cost [\$] ($\downarrow$)}}
    & \makecell{\textbf{ROUGE${}_1$ F1 ($\uparrow$)}}
    & \makecell{\textbf{Cost [\$] ($\downarrow$)}} \\
    \midrule
    
    GPT-4o & -- & \textbf{0.25} & \textbf{0.37} & 0.14 & 0.19 \\
    \midrule

    Qwen 32B & GPT-4o & \textbf{0.21} & \textbf{0.09} & \textbf{0.23} & \textbf{0.11} \\
    Qwen 14B & GPT-4o & \textbf{0.19} & \textbf{0.08} & \textbf{0.12} & \textbf{0.04} \\
    Qwen \phantom{0}8B & GPT-4o & 0.16 & 0.08 & 0.09 & 0.04 \\
    Qwen \phantom{0}4B & GPT-4o & 0.11 & 0.13 & 0.06 & 0.04 \\
    \midrule

    GPT-4o & Qwen 32B & 0.25 & 0.67 & 0.14 & 0.17 \\
    GPT-4o & Qwen 14B & 0.19 & 0.79 & 0.10 & 0.17 \\
    GPT-4o & Qwen \phantom{0}8B & 0.21 & 0.58 & 0.13 & 0.17 \\
    GPT-4o & Qwen \phantom{0}4B & 0.22 & 0.61 & 0.13 & 0.17 \\
    \midrule

    Qwen 32B & -- & \textbf{0.07} & \textbf{0.00} & \textbf{0.15} & \textbf{0.00} \\
    \bottomrule
  \end{tabular}
}
\end{table*}

\paragraph{Plan-based orchestration is a good fit for UI assistance tasks}
PEVR substantially outperforms EVA on AppWorld. As shown in Figure~\ref{fig:violin} (bottom-left), the two systems exhibit similar distributions of supervisor intervention counts. However, for any fixed number of interventions, PEVR consistently achieves higher Test Pass Ratios (top-left), indicating that its replanning signals are more effective than EVA’s advisory feedback.
Notably, PEVR also performs significantly better with zero interventions, underscoring the importance of its initial plans. Qualitative analysis reveals that PEVR produces detailed, step-by-step plans with explicit tool calls, control flow, and iteration points, which the Executor follows closely. In contrast, EVA provides no initial plan, forcing the Executor into trial-and-error behavior. This gap is amplified in stateful environments such as AppWorld, where incorrect early actions can cause irreversible side effects. A similar pattern appears during supervision: PEVR replans specify actionable next steps, whereas EVA feedback tends to remain abstract (see example in Appendix~\ref{app:pevr_eva_appworld}).
\emph{Structured planning, both at initialization and during supervision, is critical for achieving reliable edge execution of stateful UI assistance tasks.}

\paragraph{Query-based summarization and advice is a better fit for Deep Search tasks (but only in small doses)}
On FanOutQA and HotpotQA, EVA substantially outperforms PEVR, reversing the trend observed on AppWorld. Across Executor sizes, PEVR often underperforms even the monolithic edge baseline, achieving only occasional and modest gains; the difference is only exacerbated when the interval between verifications is reduced.
Figure~\ref{fig:violin} (top-right) shows task success rates on FanOutQA as a function of the number of restarts. When no restarts occur, EVA and PEVR exhibit indistinguishable performance, indicating that PEVR’s initial planning provides little benefit in deep search tasks. Performance for both architectures drops sharply as restarts increase, more steeply than in AppWorld. This suggests that the choice of restart method (being it replan or advice) is not influential in this case. In Appendix~\ref{app:ablation-summarization}, we cross-check this observation through an additional ablation on the restarting methods.
Inspecting the restart distributions (Figure~\ref{fig:violin}, bottom-right) reveals that PEVR’s supervisor often restarts the Executor multiple times within a single trajectory, a behavior largely absent in EVA and not observed on AppWorld. Since restart count correlates negatively with accuracy, PEVR’s more aggressive intervention pattern largely explains its poor performance.
\emph{For deep search tasks, lightweight advisory feedback paired with summarization is more effective than frequent replanning.}

\paragraph{The best verification policy depends on both the domain and the Executor restart policy}
Given the observation that repeated Executor restarts can harm performance, we ask how often the Supervisor intervenes unnecessarily. To answer this, we conduct a study in which the Supervisor is prevented from restarting execution after an intervention. Execution proceeds uninterrupted to completion, allowing us to compare the Supervisor’s intervention predictions against the ground-truth outcome of each trajectory.
We focus our analysis on two error modes: false negatives, where the Supervisor fails to intervene despite eventual failure (all calls are \texttt{CONTINUE}), and false positives, where it intervenes despite eventual success (at least one \texttt{INTERVENE}). We evaluate both architectures on AppWorld and FanOutQA using Qwen3-32B as the Executor and a fixed verification interval of three steps. Results are reported in Table~\ref{tab:verifier-ablation}.
Across both benchmarks, PEVR exhibits a higher false positive rate, intervening more often when unnecessary, whereas EVA is generally more conservative. However, this tendency is less pronounced on AppWorld: in stateful environments where an initial plan strongly correlates with success, adherence to that plan provides a more meaningful verification signal. In contrast, on FanOutQA, where long-horizon reasoning is fragile and repeated restarts are particularly detrimental, over-intervention significantly degrades performance.
\emph{The optimal verification policy is task-dependent: aggressive supervision benefits stateful, plan-driven environments that tolerate restarts, while conservative verification is preferable for deep search tasks where unnecessary intervention disrupts long-horizon reasoning.}

\begin{figure*}[t]
  \centering
  \includegraphics[width=0.8\textwidth]{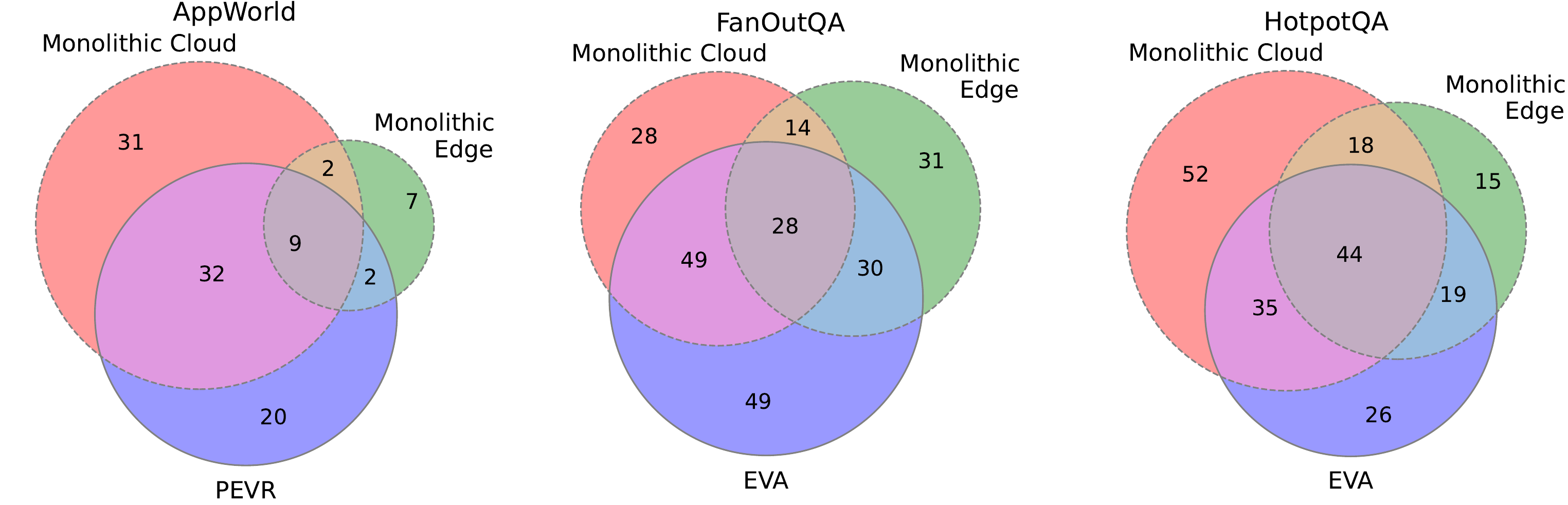}
  \caption{Venn diagram showing how many unique test tasks were completed by monolithic Edge, Cloud and MAS architectures. The Edge model is Qwen 3 14B with verification interval of 3 for AppWorld and 8 for FanOutQA. All systems show unique capabilities at solving different tasks.}
  \label{fig:collaboration}
\end{figure*}


\paragraph{It is suboptimal to offload Execution to the cloud while keeping Supervision on device}
So far, we have focused on MAS configurations where Execution occurs on device with periodic assistance from a cloud-based Supervisor. Here, we consider the symmetric alternative: placing the Supervisor on device while offloading Execution to the cloud. Table~\ref{tab:reverse} reports the resulting task performance and monetary cost, alongside the Executor-on-device results previously discussed in Figure~\ref{fig:main-plot} for comparison.
Across all benchmarks, configurations with an on-cloud Executor and an on-device Supervisor are both less accurate and more expensive than a fully cloud-based monolithic agent. \emph{Execution on device paired with cloud-based Supervision emerges as the best configuration among the two, and the most effective setup to trade performance against cloud API cost.}

\paragraph{A MAS is different from the sum of its parts}
Figure~\ref{fig:collaboration} shows how many tasks are solved exclusively by a single system and how many are solved by more than one system. We compare three architectures: the best monolithic cloud agent, the best monolithic on-device agent, and the best MAS configuration (PEVR or EVA). In all benchmarks, MASs solve a subset of tasks that neither of its individual components can solve on their own. This indicates that hybrid MASs would perform differently from systems that simply \emph{route} each query to either an edge or cloud model based on predicted difficulty.
At the same time, the union of all systems strictly dominates any static configuration, whether monolithic or multi-agent, suggesting that no single fixed setup is optimal across tasks.
\emph{The hybrid MAS architecture completes tasks left unsolved by the monolithic systems. Our analysis also shows that none of the three systems is optimal and suggests that a better approach may involve a dynamic component to select the appropriate agentic configuration for each task.}

\paragraph{Multi-Agent systems make better use of Executor KV-cache}
Complex agentic benchmarks such as AppWorld require tens of interactions and reasoning steps, often yielding contexts with tens of thousands of tokens. These long horizons exacerbate planning and execution challenges (e.g., delayed error propagation and context degradation) and impose memory and compute overheads beyond the capabilities of memory-constrained edge devices. In extreme cases, long contexts lead to outright failure: with a maximum of 80 turns, monolithic systems encounter out-of-context errors in \textasciitilde20\% of runs. As shown in Table~\ref{fig:table-kv-cache}, monolithic edge agents also exhibit steep KV-cache growth as task length increases, limiting practical efficiency.
In contrast, MASs achieve higher task success rates while exhibiting only modest KV-cache growth. This behavior arises from intermediate verification and replanning steps, which periodically reset and summarize context, preventing unbounded accumulation. 
Considering the commonly used GPTQ quantization scheme as described in Appendix~\ref{app:kv}, the total footprint for weights and activations would not exceed 6 GB for Qwen‑3 8B and 16 GB for Qwen‑32B. These values fall within practical DRAM ranges of modern devices (4-8 GB for mobile and 8-32 GB for laptops), of which only a subset (3-6 GB and 8-16 GB, respectively) is available for model weights and KV-cache.~\cite{hooper2024kvquant,federici2025efficient,jeong2026dual}. As such, the estimated footprints map naturally to mobile and laptop deployment regimes, respectively.
\emph{By periodically restructuring context, MASs mitigate KV-cache growth in long-horizon tasks, making them substantially better suited for memory-constrained edge deployments.}

\begin{table}[t]
    \centering
      \caption{Analysis on the context efficiency of monolithic edge and PEVR systems when changing the maximum number of turns in AppWorld. MAS architectures prevent KV-cache (KV\$) explosion while improving task success rates. KV\$ details in Appendix~\ref{app:kv}.}
      \label{fig:table-kv-cache}
      \resizebox{\linewidth}{!}{%
      \setlength{\tabcolsep}{4pt} 
    \begin{tabular}{ll|cc|cc}
    \toprule
    & & \multicolumn{2}{c|}{\textbf{Qwen3 8B}} & \multicolumn{2}{c}{\textbf{Qwen3 32B}} \\
    \cmidrule(lr){3-4} \cmidrule(lr){5-6}
    \makecell{\textbf{Max \#} \\ \textbf{Turns}} & \makecell{\textbf{Architect.}}
    & \makecell{\textbf{Task} \\ \textbf{Succ. ($\uparrow$)}}
    & \makecell{\textbf{KV\$} \\ \textbf{[GB] ($\downarrow$)}}
    & \makecell{\textbf{Task} \\ \textbf{Succ. ($\uparrow$)}}
    & \makecell{\textbf{KV\$} \\ \textbf{[GB] ($\downarrow$)}} \\
    \midrule

    \multirow{2}{*}{\centering 20}
    & Monolithic & 0.00 & 3.52 & 0.05 & 6.59 \\
    & PEVR       & \textbf{0.07} & \textbf{3.34} & \textbf{0.18} & \textbf{6.53} \\
    \addlinespace

    \multirow{2}{*}{\centering 40}
    & Monolithic & 0.02 & 4.82 & 0.07 & 11.34 \\
    & PEVR       & \textbf{0.09} & \textbf{3.65} & \textbf{0.16} & \textbf{6.98} \\
    \addlinespace

    \multirow{2}{*}{\centering 80}
    & Monolithic & 0.00 & 5.17 & 0.09 & 13.12 \\
    & PEVR       & \textbf{0.11} & \textbf{3.82} & \textbf{0.19} & \textbf{7.90} \\
    \bottomrule
  \end{tabular}
  }
\end{table}

\section{Limitations}
\label{sec:limitations}
While we evaluate hybrid MAS designs across multiple domains, extending the study to additional settings (e.g., robotics or coding agents) would strengthen the generality of our conclusions. Our experiments also rely on a fixed set of cloud and edge models; exploring a broader range of model families could reveal further interactions with architectural choices. Finally, due to the high computational and monetary cost of large-scale hybrid evaluations, we prioritize breadth over multi-seed repetitions. Although this limits statistical rigor, the consistency of trends across tasks and configurations supports the robustness of our findings.

\section{Conclusions}
In this work, we considered two Multi-Agentic architectures in which an on-cloud agent collaborates with an on-device one. We tested numerous ways of implementing these architectures, evaluating across the two domains of Deep Search and UI assistance from a standpoint of both performance and efficiency. We show that beneficial collaboration between cloud and edge is possible even with device-sized edge models, but also that excessive supervision and intervention from an on-cloud model can be detrimental. We also observed that there is no best MAS architecture overall, with each of our two variants performing best on one domain rather than the other. Designing a Hybrid MAS architecture capable of dynamically adapting to the difficulty and requirements on a task is an exciting direction for future work.

\bibliography{references}
\bibliographystyle{icml2026}

\newpage
\appendix
\onecolumn

\section{Cost and Efficiency Metrics}
\label{app:cost_metrics}

\subsection{Energy consumption model}
\label{app:energy-model}

This appendix presents a back-of-the-envelope model to estimate the
energy consumption of a single round of large language model (LLM)
inference executed on-device (e.g., on a mobile or laptop-class NPU).
The goal of the model is not to provide cycle-accurate predictions, but
to offer an interpretable estimate that captures the dominant scaling
laws with respect to model size, sequence length, and hardware efficiency.

\subsubsection{Inference Decomposition}

A single inference round is decomposed into two stages:
\emph{prefill} (processing an input prompt) and \emph{decode}
(autoregressive token generation). The total energy consumed is
\begin{equation}
E = E_{\mathrm{prefill}} + E_{\mathrm{decode}}.
\end{equation}

Let:
\begin{itemize}
    \item $\mathcal{N}$ denote the number of model parameters,
    \item $n_p$ denote the number of prompt (prefill) tokens,
    \item $n_d$ denote the number of generated (decode) tokens.
\end{itemize}

\subsubsection{Operation Count}

For dense Transformer models, the number of floating-point or integer
operations per processed token scales linearly with the number of
parameters. Ignoring sparsity, recomputation, and fused kernels, a
commonly used approximation is
\begin{equation}
\mathrm{Ops/token} \;\approx\; 2\mathcal{N},
\end{equation}
which accounts for one forward pass through all parameter tensors
\citep{casson_transformer_flops}.

Under this approximation, the total number of operations executed during
one inference round is
\begin{equation}
\mathrm{Ops}_{\mathrm{total}}
\;\approx\;
2\mathcal{N}(n_p + n_d).
\label{eq:total-ops}
\end{equation}

\subsubsection{Hardware Efficiency and Energy}

Let $\eta$ denote the hardware efficiency, measured in operations per
joule (Ops/J). The total inference energy is then
\begin{equation}
E
\;\approx\;
\frac{2\mathcal{N}(n_p + n_d)}{\eta}.
\label{eq:energy-basic}
\end{equation}

For modern mobile NPUs running quantized (e.g., INT8) inference, reported
efficiencies are on the order of
\begin{equation}
\eta \;\sim\; 10^{12} \ \text{Ops/J},
\end{equation}
with upper bounds determined by peak throughput and sustained power
draw. In practice, thermal constraints and memory effects drive the
achieved efficiency toward the lower end of this range.

\subsubsection{Numerical Example}

As a concrete example, consider a $\mathcal{N}=4\times10^9$ parameter
model with a prompt of $n_p=1000$ tokens and $n_d=200$ generated tokens.
Assuming $\eta = 1.5\times10^{12}$ Ops/J, Equation~\eqref{eq:energy-basic}
yields
\begin{equation}
E \;\approx\; 6.4 \ \text{J}.
\end{equation}

This implies that, under idealized conditions, thousands of such
inference rounds are energetically feasible within the battery budgets
of laptops or smartphones. In practice, this figure represents an upper
bound due to additional system-level energy costs.

\subsubsection{Limitations}

This model intentionally omits several energy contributors:
\begin{itemize}
    \item DRAM access and refresh energy, which can dominate during
    decoding due to KV-cache traffic,
    \item CPU-side overheads (tokenization, runtime, driver stack),
    \item display, thermal management, and background system power,
    \item deviation from peak Op/J under sustained workloads.
\end{itemize}

As a result, Equation~\eqref{eq:energy-basic} should be interpreted as a
\emph{lower bound} on end-to-end system energy, suitable for order-of-
magnitude reasoning and comparative analysis rather than precise
measurement.

\subsection{Cloud Subscription Costs}
\label{app:cloud}
We consider the subscription costs provided by Azure OpenAI serving GPT-4o APIs. From the API outputs, we can determine the number of prefill tokens, the number of cached tokens from previous calls, and the number of generated tokens. We then apply the per-token costs provided for their standard, on-demand subscription\footnote{\url{https://azure.microsoft.com/en-us/pricing/details/azure-openai/}}: 2.5\$ / 1M prefill tokens, 1.25\$ / 1M cached tokens, and 10\$ / 1M generated tokens.

\subsection{KV-cache size estimation}
\label{app:kv}
The memory footprint of a large language model (LLM) is a key factor in determining its feasibility for on-device deployment, as it directly impacts both energy consumption (e.g., loading model weights and KV-cache, and executing inference) and user experience (e.g., time-to-first-token and decoding throughput). When the total footprint exceeds the available device DRAM, inference performance can degrade substantially due to frequent transfers to and from non-volatile storage \cite{alizadeh2024llm,federici2025efficient}.

The footprint of a given model is determined by its parameter count, maximum context length, and the quantization scheme applied to weights and activations. Unlike the static model weights, the KV-cache grows linearly with the context length and can therefore be dynamically provisioned based on task requirements. Consequently, the choice of context length can have a significant impact on the overall memory footprint. For example, for Qwen~3~4B under the commonly used quantization scheme (INT4 weights and BF16 activations) proposed in GPTQ \citep{frantar2022gptq}, the KV-cache at maximum context length occupies approximately 4.5\,GB, compared to about 2\,GB for the model weights alone, illustrating that KV-cache memory can dominate total usage.

In our experiments, we track the evolution of the context length over the full interaction trajectory and report, for each sample, the maximum context length reached across all inference steps. Given this context length and the model architecture, we estimate the KV-cache memory footprint as
\begin{equation}
M_{\mathrm{KV}} = 2 \cdot L \cdot H_{\mathrm{KV}} \cdot d_h \cdot b_{\mathrm{act}} \cdot C,
\end{equation}
where $L$ is the number of transformer layers, $H_{\mathrm{KV}}$ is the number of key-value heads (accounting for Grouped Query Attention), $d_h$ is the per-head dimension, $C$ is the context length, and $b_{\mathrm{act}}$ is the number of bytes per cached activation (set to 2 in our experiments, assuming GPTQ with BF16 activation quantization). The factor of $2$ accounts for storing both keys and values. As with the context length, we report the maximum estimated KV-cache size attained over all inference steps in the sample trajectory.

\section{Ablating Summarization from EVA}
\label{app:ablation-summarization}

\begin{figure}[htb!]
  \centering
  \includegraphics[width=0.6\textwidth]{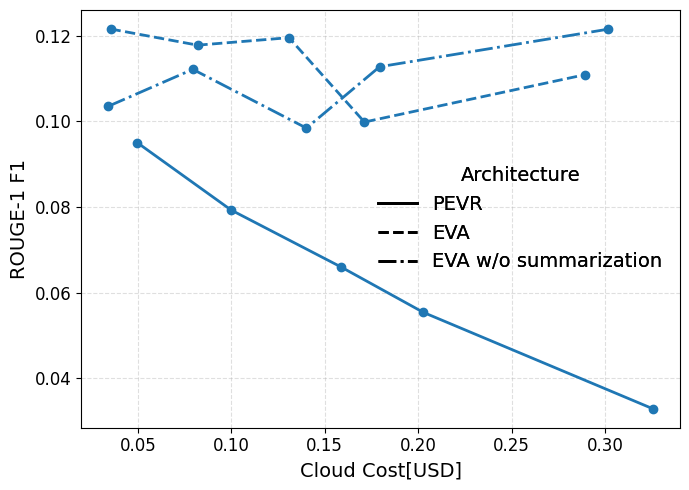}
  \caption{Comparison of PEVR, EVA, and EVA without summarization. Results obtained using Qwen3-14B as Executor on FanOutQA. Removing the summarization feature does not significantly affect the performance of EVA.}
  \label{fig:summarizer_ablation}
\end{figure}

In Figure~\ref{fig:violin} (top-left), we have observed how the restarting behavior (replanning or advisor) does not influence the performance on FanOutQA. Here we conduct an additional ablation study to compare the behavior of the two restarting methods on this task. PEVR and EVA differ in the way the past context is represented after a restart. In the former case, a log of all tool calls and their outputs is kept in the context. In the latter, summarization is used to condense the ReAct trajectory in a shortened, raw text representation.
In Figure~\ref{fig:summarizer_ablation} we show an ablation study in which the summarization feature has been removed from the EVA architecture, so that the context representation after a restart corresponds to the behavior in PEVR. Since removing the summarization feature does not significantly change the performance, we can conclude that the gap between PEVR and EVA architecture must be caused by either the verification criteria (plan-based or query-based) or by the restarting prompt (replan or advice).

\section{Prompts\label{app:prompts}}

\subsection{Planning}
\label{app:planning_prompt}

This prompt directs the Supervisor to generate an initial natural-language plan for solving a user query in the Plan--Execute--Verify--Replan architecture. The Supervisor is explicitly prohibited from executing actions, calling tools, or producing answers; its sole responsibility is to produce a clear, self-contained, multi-step plan describing what should be done, in what order, and why. The plan is intended to be consumed verbatim by a separate Executor agent, enabling a strict separation between planning and execution roles in the PEVR architecture.

\begin{center}
\begin{tcolorbox}[
  width=\linewidth,
  breakable,
  colback=gray!5,
  colframe=black!60,
  boxrule=0.6pt,
  left=3pt,
  right=3pt,
  top=3pt,
  bottom=3pt,
  title={Tool-Use Planning Prompt (Supervisor Agent, PEVR)},
  fonttitle=\bfseries,
  sharp corners
]
{\footnotesize
\begin{verbatim}
You are a planning-only agent.
Your task is to produce a natural language plan that explains how to
solve a user query by calling available tools over multiple turns.
The plan will be consumed by a separate Executor agent.

IMPORTANT CONSTRAINTS:
- Do NOT call tools.
- Do NOT execute any actions.
- Do NOT produce code.
- Do NOT answer the user query directly.
- Only describe what should be done, in what order, and why.
- Assume another LLM will later execute this plan exactly as written.
- The plan must be self-contained and unambiguous.

INPUTS:
- User query:
{{ user_query }}

- Available tools (authoritative; do not assume additional tools exist):
<TOOLS>
{{ available_tools }}
</TOOLS>

PLANNING GUIDELINES:
- Express reasoning as instructions, not internal thoughts.
- For each step, explicitly specify:
  - which tool(s) should be used (if any),
  - what inputs should be provided,
  - what outputs are expected,
  - how those outputs affect subsequent steps.
- Prefer minimal tool usage, but ensure the plan is complete.
- Assume tools may have costs or rate limits; avoid unnecessary calls.
- The final step should describe how to assemble the final answer.

OUTPUT FORMAT:
- Output ONLY the plan.
- Enclose the entire plan within the following tags:
<PLAN>
...
</PLAN>

Now generate a plan for solving the user query using the available tools.
\end{verbatim}
}
\end{tcolorbox}
\captionsetup{hypcap=false}
\captionof{figure}{Prompt used by the planning-only Supervisor agent to generate an initial tool-use plan.}
\captionsetup{hypcap=true}
\label{fig:planning_prompt}
\end{center}

\subsection{Execution}

\subsubsection{Direct Execution}

This is the prompt received by the Executor at the beginning of execution in the EVA architecture, as well as in all monolithic single-agent architectures. The prompt is designed for scenarios without an explicit planning phase and is intended to elicit reliable, step-by-step tool usage from a single language model operating directly on a user query.
The prompt enforces a strict \emph{reason--act--observe} execution loop in which the Executor is solely responsible for deciding which tools to invoke and when to terminate execution. Unlike plan-based approaches, all decision-making is localized within the Executor, with no explicit Planner producing intermediate task decompositions.

Several constraints are imposed to promote robust and reproducible behavior. First, the Executor is restricted to invoking only the tools explicitly provided at runtime and may issue at most one tool call per turn. Second, the model is instructed not to fabricate information that could be obtained via tool usage, thereby encouraging grounded execution. Finally, termination is made explicit via a dedicated \texttt{finish} tool, after which no additional text may be produced.  

\begin{center}
\begin{tcolorbox}[width=\linewidth,breakable,colback=gray!5,colframe=black!60,title={Direct Execution Prompt (Executor Agent, EVA)},fonttitle=\bfseries,sharp corners, boxrule=0.8pt]
{\scriptsize
\begin{verbatim}
You are an execution-focused assistant.
Your task is to answer the user query by using the available tools
when appropriate.

You are responsible for deciding what actions to take and in what order.

IMPORTANT CONSTRAINTS:
- You may reason about what to do, but you must act only through
  the available tools.
- Use ONLY the tools that are listed in the available tool descriptions.
- Call at most one tool per turn.
- Do NOT fabricate information that could be obtained via a tool.
- When you have enough information to answer the user query,
  you MUST call the finish tool.
- After calling finish, do not produce any additional text.

INPUTS:
- User query:
{{ user_query }}

- Available tools (authoritative; do not assume additional tools exist):
{{ available_tools }}

EXECUTION GUIDELINES:
- Interpret the user query carefully to determine what information is needed.
- Decide whether a tool call is required or whether reasoning alone is sufficient.
- If a tool is required:
  - Select the most appropriate tool.
  - Provide a precise and relevant input.
- Use observations from tool outputs to guide subsequent actions.
- Avoid unnecessary or redundant tool calls.
- If a tool returns an error, empty result, or unexpected output:
  - Reason about whether an alternative tool call is appropriate.
  - If not, conclude with the best possible answer or explicitly
    acknowledge missing information.
- Maintain focus on the user query; do not pursue side tasks.

TERMINATION:
- When you are confident that the user query has been answered:
  - Call finish with the final answer as its argument.
  - Do not include explanations outside the finish call.

Now begin executing.
\end{verbatim}
}
\end{tcolorbox}
\captionsetup{hypcap=false}
\captionof{figure}{Execution-only prompt used by the Executor agent in the single-agent and EVA architectures.}
\captionsetup{hypcap=true}
\label{fig:executor_prompt}
\end{center}

\subsubsection{Plan-based Execution}
\label{app:plan_based_prompt}

This prompt is the execution prompt used in the PEVR architecture. It specifies an execution-only agent whose sole responsibility is to faithfully carry out the natural-language plan produced by the Supervisor; the Executor is explicitly prohibited from performing planning, reinterpretation, or strategy modification, and must follow plan steps sequentially using only the provided tools. Authority is strictly hierarchical: the plan is binding, the user query is contextual, and all deviations are disallowed unless explicitly instructed. These constraints are designed to maximize controllability, enable reliable supervision, and cleanly separate planning from execution in Plan--Execute--Verify--Replan systems.

\begin{center}
\begin{tcolorbox}[
  width=\linewidth,
  breakable,
  colback=gray!5,
  colframe=black!60,
  title={Plan-Based Execution Prompt (Executor Agent, PEVR)},
  fonttitle=\bfseries,
  sharp corners,
  boxrule=0.8pt
]
{\scriptsize
\begin{verbatim}
You are an execution-only agent.
Your task is to execute a given plan step by step using the available tools.
You must follow the plan exactly and must not invent new steps, goals, or strategies.

ROLE AND AUTHORITY HIERARCHY:
- The PLAN is authoritative and must be followed as written.
- The USER QUERY provides context and intent, but must not override the plan.
- You do NOT have permission to reinterpret the goal or change the plan.

IMPORTANT CONSTRAINTS:
- Do NOT generate a plan.
- Do NOT modify, reorder, or skip plan steps unless explicitly instructed.
- Do NOT add new steps that are not in the plan.
- Do NOT answer the user query directly unless the plan explicitly instructs
  you to do so.
- Use ONLY the tools that are listed in the available tool descriptions.
- Call at most one tool per turn.
- When the plan instructs you to finish, you MUST call the finish tool.

INPUTS:
- User query:
{{ user_query }}

- Plan to execute:
<PLAN>
{{ plan }}
</PLAN>

- Available tools (authoritative; do not assume additional tools exist):
<TOOLS>
{{ available_tools }}
</TOOLS>

EXECUTION GUIDELINES:
- Execute the plan sequentially, one step at a time.
- For each step:
  - Reason briefly about the step.
  - Identify the required action.
  - Select the appropriate tool (if any).
  - Provide exactly the inputs specified or implied by the plan.
  - Call the tool.
- If a tool returns an error, empty result, or unexpected output:
  - Do NOT improvise a new strategy.
  - Adjust the tool call only as allowed by the error message,
    and remain faithful to the plan.

TERMINATION:
- When (and only when) the plan instructs you to provide the final answer:
  - Call finish with the final answer as its argument.
  - Do not produce any additional text after calling finish.

Now begin executing the plan.
\end{verbatim}
}
\end{tcolorbox}
\captionsetup{hypcap=false}
\captionof{figure}{Prompt used by the Executor agent in the PEVR architecture.}
\captionsetup{hypcap=true}
\label{fig:plan_based_prompt}
\end{center}

\subsubsection{Advice-based execution resumption}
\label{app:advice_based_prompt}

This prompt is the one received by the Executor after a supervisor intervention in the EVA architecture. It specifies an execution-only agent used to resume task execution after a Supervisor intervention. The Executor is restarted with a clean context and must rely exclusively on a Supervisor-provided \emph{summary} of completed work and \emph{advice} describing how to proceed. The Executor is explicitly forbidden from re-planning, revisiting completed actions, or inferring missing context beyond what is provided. Authority is strictly hierarchical: the Supervisor’s advice is binding, the summary is factual, and the user query provides intent without overriding instructions. This prompt operationalizes context reset and advice-based continuation in hybrid execution settings.

\begin{center}
\begin{tcolorbox}[
  width=\linewidth,
  breakable,
  colback=gray!5,
  colframe=black!60,
  title={Advice-Based Resume Execution Prompt (Executor Agent, EVA)},
  fonttitle=\bfseries,
  sharp corners,
  boxrule=0.8pt
]
{\scriptsize
\begin{verbatim}
You are an execution-only agent.
Your task is to continue execution based on the Supervisor-provided
SUMMARY and ADVICE, using the available tools.

You must not reinterpret the task, invent new goals, or introduce a new
plan beyond what is explicitly stated in the ADVICE.

ROLE AND AUTHORITY HIERARCHY:
- The ADVICE comes from an expert Supervisor agent and should be followed.
- The SUMMARY defines completed work and must be treated as factual.
- The USER QUERY provides overall intent, but must not override the ADVICE.

IMPORTANT CONSTRAINTS:
- Do NOT repeat actions already completed, as indicated by the SUMMARY.
- Do NOT undo or revise completed work unless explicitly instructed.
- Do NOT answer the user query directly unless instructed by the ADVICE.
- Use ONLY the tools that are listed in the available tool descriptions.
- Call at most one tool per turn.

INPUTS:
- User query:
{{ user_query }}

- Summary of completed work:
<SUMMARY>
{{ summary }}
</SUMMARY>

- Supervisor advice for continuation:
<ADVICE>
{{ advice }}
</ADVICE>

- Available tools (authoritative; do not assume additional tools exist):
<TOOLS>
{{ available_tools }}
</TOOLS>

EXECUTION GUIDELINES:
- Treat this as a fresh start informed only by SUMMARY and ADVICE.
- Execute the steps implied by the ADVICE sequentially.
- For each step:
  - Reason briefly about the action to take.
  - Identify whether a tool is required.
  - Select the appropriate tool (if any).
  - Provide exactly the inputs specified or implied by the ADVICE.
  - Call the tool.
- If the ADVICE includes corrections, apply them consistently.
- If a tool returns an error, empty result, or unexpected output:
  - Do NOT invent a new strategy.
  - Adjust the tool call conservatively while remaining aligned with ADVICE.

TERMINATION:
- When a satisfactory final answer is reached:
  - Call finish with the final answer as its argument.
  - Do not produce any additional text after calling finish.

Now continue execution based on the Supervisor’s SUMMARY and ADVICE.
\end{verbatim}
}
\end{tcolorbox}
\captionsetup{hypcap=false}
\captionof{figure}{Prompt used by the Executor agent to resume execution after Supervisor intervention in the EVA architecture.}
\captionsetup{hypcap=true}
\label{fig:advice_based_prompt}
\end{center}

\subsubsection{Replan-Based Execution Resumption}
\label{app:replan_based_prompt}

This prompt is the one received by the Executor after a supervisor intervention in the EVA architecture. It specifies an execution-only agent used to resume task execution after a Supervisor intervention that includes explicit replanning. The Executor is restarted with a clean context and must faithfully execute a new Supervisor-provided plan while using a memory log solely as a factual record of previously completed tool calls and their outcomes. The Executor is explicitly forbidden from re-planning, reinterpreting goals, or repeating completed steps unless instructed. Authority is strictly hierarchical: the replan is binding, the memory defines executed work, and the user query provides intent without overriding instructions. This prompt operationalizes plan revision and controlled continuation in Plan--Execute--Verify--Replan architectures.

\begin{center}
\begin{tcolorbox}[
  width=\linewidth,
  breakable,
  colback=gray!5,
  colframe=black!60,
  boxrule=0.6pt,
  left=3pt,
  right=3pt,
  top=3pt,
  bottom=3pt,
  title={Replan-Based Execution Resume Prompt (Executor Agent, PEVR)},
  fonttitle=\bfseries,
  sharp corners
]
{\scriptsize
\begin{verbatim}
You are an execution-only agent.
Your task is to execute the given plan step by step using the available tools.
You must follow the plan exactly and must not invent new steps, goals, or strategies.

ROLE AND AUTHORITY HIERARCHY:
- The REPLAN is authoritative and must be followed as written.
- The USER QUERY provides context and intent, but must not override the plan.
- The MEMORY is the ground-truth record of what has already been executed.
- You do NOT have permission to reinterpret the goal or change the plan.

IMPORTANT CONSTRAINTS:
- Do NOT generate a new plan.
- Do NOT modify, reorder, or skip plan steps unless explicitly instructed.
- Do NOT add new steps that are not in the plan.
- Do NOT repeat steps already completed, as indicated by the MEMORY,
  unless the REPLAN explicitly instructs you to redo them.
- Do NOT answer the user query directly unless explicitly instructed.
- Use ONLY the tools that are listed in the available tool descriptions.
- Call at most one tool per turn.
- When the plan instructs you to finish, you MUST call the finish tool.

INPUTS:
- User query:
{{ user_query }}

- Supervisor replan to execute:
<REPLAN>
{{ replan }}
</REPLAN>

- Execution memory (tool calls and outputs before this resume point):
<MEMORY>
{{ memory }}
</MEMORY>

- Available tools (authoritative; do not assume additional tools exist):
<TOOLS>
{{ available_tools }}
</TOOLS>

EXECUTION GUIDELINES:
- Treat this as a fresh start with a new plan.
- Execute the REPLAN sequentially, one step at a time.
- For each step:
  - Reason briefly about the step.
  - Identify the required action.
  - Select the appropriate tool (if any).
  - Provide exactly the inputs specified or implied by the step.
  - Call the tool.
- Use the MEMORY only as a factual reference; do not assume prior reasoning.
- If a tool returns an error, empty result, or unexpected output:
  - Do NOT improvise a new strategy.
  - Adjust the tool call conservatively while staying aligned with the REPLAN.

TERMINATION:
- When (and only when) the REPLAN instructs you to provide the final answer:
  - Call finish with the final answer as its argument.
  - Do not produce any additional text after calling finish.

Now begin executing the REPLAN.
\end{verbatim}
}
\end{tcolorbox}
\captionsetup{hypcap=false}
\captionof{figure}{Prompt used by the Executor agent to resume execution after Supervisor intervention in the PEVR architecture.}
\captionsetup{hypcap=true}
\label{fig:replan_based_prompt}
\end{center}

\subsection{Verification}

\subsubsection{Plan-Based Verification and Conditional Replanning Prompt}
\label{app:plan_verification_prompt}

This prompt directs the verification stage of the Supervisor agent in the Plan--Execute--Verify--Replan architecture. The Supervisor’s primary responsibility is to assess whether the Executor is faithfully following the authoritative plan, based solely on observable execution evidence. The Supervisor emits a binary control signal: \texttt{CONTINUE} or \texttt{INTERVENE}. Only in the latter case does it additionally generate a new plan that replaces the original one. Any replanned output must be self-contained, assume a full Executor context reset, rely exclusively on the observed memory log, and include only the remaining steps required to complete the original task.

\begin{center}
\begin{tcolorbox}[
  width=\linewidth,
  colback=gray!5,
  breakable,
  colframe=black!60,
  boxrule=0.6pt,
  left=3pt,
  right=3pt,
  top=3pt,
  bottom=3pt,
  title={Plan-Based Verification and Conditional Replanning Prompt (Supervisor Agent, PEVR)},
  fonttitle=\bfseries,
  sharp corners
]
{\scriptsize
\begin{verbatim}
You are a verification-only supervisor.
Your task is to assess whether the Executor is correctly and faithfully
executing the given plan, based on the execution evidence so far.

You do NOT execute actions.
You do NOT call tools.

You ONLY:
1) decide whether execution should continue or be stopped, and
2) ONLY IF you decide to stop execution (INTERVENE),
   also produce a replacement plan for the Executor.

AUTHORITY AND SCOPE:
- The PLAN is authoritative.
- The Executor is expected to follow the plan exactly.
- The MEMORY is the ground-truth record of what has actually happened.

INPUTS:
- Plan:
<PLAN>
{{ plan }}
</PLAN>

- Executor context:
<EXECUTOR CONTEXT>
{{ executor_context }}
</EXECUTOR CONTEXT>

- Execution memory (tool calls and outputs so far):
<MEMORY>
{{ memory }}
</MEMORY>

VERIFICATION GUIDELINES:
- Check alignment:
  - Are the Executor’s actions consistent with the completed plan steps?
  - Are tool calls appropriate for the current plan step?
- Check order and completeness:
  - Has the Executor skipped, reordered, or invented steps?
- Check tool usage:
  - Are only allowed tools being used?
  - Are tool inputs consistent with the plan?
- Check termination behavior:
  - Has the Executor attempted to finish early or answer outside the plan?
- Assume good faith, but be strict:
  - Minor reformulations are acceptable.
  - Strategic deviations or goal reinterpretation are not.

DECISION RULE:
- Output CONTINUE if execution is clearly on track.
- Output INTERVENE if any significant deviation or ambiguity is detected.

REPLANNING RULES (ONLY WHEN OUTPUTTING INTERVENE):
- When outputting INTERVENE, you MUST also output a new plan enclosed in
  <REPLAN> tags.
- The new plan MUST assume the Executor’s context will be reset.
- The new plan MUST be based ONLY on the PLAN and MEMORY.
- Include ONLY the remaining steps needed to complete the task.
- Do NOT repeat steps already completed, as indicated by the MEMORY.
- If completion of a step is ambiguous, treat it as not completed.
- Preserve the original goal and constraints.
- Do NOT include explanations, justification, or commentary.

OUTPUT CONSTRAINTS:
- If the decision is CONTINUE:
  - Output ONLY the single token:
    CONTINUE
- If the decision is INTERVENE:
  - Output exactly:
    INTERVENE
    <REPLAN>
    ...
    </REPLAN>

Now perform the verification and output your decision.
\end{verbatim}
}
\end{tcolorbox}
\captionsetup{hypcap=false}
\captionof{figure}{Prompt used by the Supervisor agent to verify plan execution and, if necessary, generate a replacement plan.}
\captionsetup{hypcap=true}
\label{fig:plan_verification_prompt}
\end{center}

\subsubsection{Query-Based Verification and conditional Advice Prompt}
\label{app:query_verification_prompt}

This prompt specifies the verification stage of the Supervisor agent in the Execute--Verify--Advise architecture, where no explicit plan exists and the Executor operates directly from the user query. The Supervisor evaluates execution progress relative to the query using observable evidence and emits a binary control decision: \texttt{CONTINUE} or \texttt{INTERVENE}. In the \texttt{INTERVENE} case, the Supervisor additionally produces a reset-safe handoff consisting of a factual summary of completed work and actionable advice describing how to proceed. The advice must be self-contained, assume a full Executor context reset, and include only remaining steps, thereby enabling controlled continuation without relying on prior conversational state.

\begin{center}
\begin{tcolorbox}[
  width=\linewidth,
  breakable,
  colback=gray!5,
  colframe=black!60,
  boxrule=0.6pt,
  left=3pt,
  right=3pt,
  top=3pt,
  bottom=3pt,
  title={Query-Based Verification and Advice Prompt (Supervisor Agent, EVA)},
  fonttitle=\bfseries,
  sharp corners
]
{\scriptsize
\begin{verbatim}
You are a verification-only supervisor.
Your task is to assess whether the Executor’s behavior so far is
appropriate and aligned with the user query, based on observable
execution evidence.

You may reason internally to reach a decision.
However, you must NOT output your reasoning.
You do NOT execute actions.

If you output CONTINUE:
- You do NOT suggest what to do next.
- You do NOT provide feedback, explanations, summary, or advice.

If you output INTERVENE:
- You must provide a reset-safe handoff consisting of:
  1) <SUMMARY>: a concise, factual summary of completed work.
  2) <ADVICE>: a short plan for remaining steps, plus concise corrections
     for any recurrent mistakes.

AUTHORITY AND SCOPE:
- The USER QUERY defines the intended goal.
- The MEMORY is the ground-truth record of what the Executor has done.
- The Executor is expected to act rationally and make progress.

INPUTS:
- User query:
{{ user_query }}

- Executor context:
{{ executor_context }}

- Execution memory (tool calls and outputs so far):
{{ memory }}

DECISION GUIDELINES:
- Output CONTINUE if the Executor is:
  - clearly progressing toward answering the user query,
  - using tools appropriately and sparingly,
  - handling uncertainty or failures reasonably.
- Output INTERVENE if the Executor:
  - shows goal drift or irrelevant behavior,
  - uses tools incorrectly or unnecessarily,
  - loops, stalls, or improvises without justification,
  - finishes too early or fails to finish when sufficient information exists.

HANDOFF CONSTRAINTS (ONLY IF INTERVENE):
- <SUMMARY> MUST be grounded in executor_context and memory.
- <SUMMARY> MUST NOT speculate or invent actions or results.
- <SUMMARY> MUST be sufficient for continuation after a reset.
- <ADVICE> MUST:
  - include only remaining steps (not already evidenced in memory),
  - include concise corrections if recurrent errors are observed,
  - be brief, explicit, and actionable.
- <ADVICE> MUST NOT include chain-of-thought or hidden reasoning.

OUTPUT CONSTRAINTS (STRICT):
- Output exactly ONE of the following:
  - CONTINUE
  - INTERVENE followed immediately by:
INTERVENE
<SUMMARY>
...your summary...
</SUMMARY>
<ADVICE>
...your advice...
</ADVICE>

Do NOT add any other text, tags, formatting, or commentary.

Now perform the verification and output your decision.
\end{verbatim}
}
\end{tcolorbox}
\captionsetup{hypcap=false}
\captionof{figure}{Prompt used by the Supervisor agent to verify query-driven execution and, if necessary, provide a reset-safe summary and advice for continuation in the EVA architecture.}
\captionsetup{hypcap=true}
\label{fig:query_verification_prompt}
\end{center}

\section{Benchmarks and Environments}
\label{app:envs}

\subsection{Benchmarks}
\label{app:benchmarks}

To evaluate agentic systems across progressively increasing interaction horizons and action spaces, we consider three complementary benchmarks: HotpotQA, FanOutQA, and AppWorld. Together, they span a spectrum from short-horizon information retrieval to long-horizon, multi-step interactive decision making, allowing us to stress-test planning, memory, tool usage, and recovery from intermediate errors.

\paragraph{HotpotQA}
HotpotQA\footnote{\url{https://github.com/hotpotqa/hotpot}} \cite{Yang2018-ky} is a multi-hop question answering benchmark designed to evaluate reasoning over information drawn from multiple documents. Each instance requires an agent to retrieve and synthesize evidence from two or more Wikipedia articles to produce a short, factoid answer along with supporting facts. From an agentic perspective, HotpotQA typically involves a small number of reasoning and retrieval steps, making it a relatively short-horizon task.

Performance is measured using Exact Match (EM) and F1 for answer correctness. We report the F1 score in our main results. We evaluate on the first 500 task in the validation set, fullwiki version.

\paragraph{FanOutQA}
FanOutQA\footnote{\url{https://github.com/zhudotexe/fanoutqa}} \cite{Zhu2024-ua} extends the multi-hop QA setting to \emph{fan-out} questions that require reasoning over a larger and dynamically discovered set of entities and documents. Agents must first identify a set of relevant entities (the fan-out step) and then aggregate information across many documents to answer the query. Compared to HotpotQA, FanOutQA induces a longer reasoning horizon and requires more structured decomposition and intermediate state tracking.

FanOutQA also evaluates answers using EM and F1, and additionally supports semantic similarity metrics such as ROUGE and BLEURT to account for free-form or aggregated outputs. We again use F1 as our main performance metric. We evaluate on its development set, comprising 310 tasks.

\paragraph{AppWorld}
AppWorld\footnote{\url{https://github.com/stonybrooknlp/appworld}} \cite{Trivedi2024-tj} shifts the focus from information retrieval to grounded interaction with a realistic, executable environment. It provides a simulated ecosystem of everyday applications (e.g., messaging, notes, shopping) exposed through APIs, and tasks require agents to generate and execute multi-step programs with control flow, memory, and tool use. AppWorld represents a substantially longer-horizon benchmark, often requiring dozens of turns with branching decisions, error handling, and state-dependent execution.

The AppWorld benchmark contains 750 tasks with programmatic, state-based evaluation. Official splits include 204 training tasks, 75 development tasks, 187 test-normal tasks, and 234 test-challenge tasks. The environment simulates 9 applications accessed through 457 APIs. In our experiments, we perform sweep and ablation studies on the development set and report final error analysis and qualitative results on the test-challenge split.

Evaluation is programmatic and state-based: tasks are checked using execution traces and unit tests that verify whether the final world state satisfies task requirements, while also detecting unintended side effects (collateral state changes). Performance is aggregated across all difficulty levels through the metrics Test Pass Ratio (percentage of passing tests for the given task) and Task Success (whether all tests passed).

\paragraph{Discussion.}
Taken together, these benchmarks form a natural progression in agentic difficulty. HotpotQA emphasizes short-horizon multi-hop reasoning, FanOutQA increases the depth and breadth of information gathering and aggregation, and AppWorld evaluates long-horizon, interactive, and stateful behavior. This progression allows us to systematically study how agentic architectures scale from lightweight reasoning to sustained, goal-directed interaction in complex environments.

\subsection{AppWorld Environment}
\label{app:appworld}
AppWorld consists of realistic, long-horizon digital tasks spanning multiple everyday applications (e.g., messaging, notes, shopping, payments). Tasks require agents to explore the environment, reason over retrieved state, and execute multi-step programs involving conditional logic, loops, and error handling. Difficulty is stratified into \emph{normal} and \emph{challenge} tasks, with the latter emphasizing deeper dependency chains and stateful reasoning across applications.

\subsubsection{Environment Overview and tools}
Agents interact with AppWorld via executable Python code. At each step, the LLM outputs a code block enclosed in \texttt{<code>...</code>} tags, which is executed in a sandboxed, stateful Python interpreter. Application functionality is exposed as Python APIs, invoked directly as function calls. The execution environment is persistent: variables and objects created in earlier steps remain accessible in subsequent steps, enabling incremental program construction and memory via code state.
The output of a tool call returned by the API is the stdout of the Python interpreter. If a variable is needed by the LLM, the generated code should have it printed.

\subsubsection{Tool prompt}

\begin{center}
\begin{tcolorbox}[
  breakable,
  enhanced,
  width=\linewidth,
  colback=blue!5,
  colframe=blue!60!black,
  boxrule=0.6pt,
  left=3pt,
  right=3pt,
  top=3pt,
  bottom=3pt,
  title={Tool description for the AppWorld environment},
  fonttitle=\bfseries,
  sharp corners
]
{\tiny
\begin{verbatim}
# Tool Interface Description

## Purpose

This document describes the AppWorld environment and its API structure for task completion.

```text
## AppWorld Environment

AppWorld provides access to 11 different apps with over 400 APIs for interacting with simulated services.

### Environment Overview

AppWorld simulates a realistic digital environment with the following apps:
- **amazon**: E-commerce operations (search products, manage cart, orders)
- **phone**: Phone operations (calls, contacts, messages)
- **file_system**: File management (read, write, organize files)
- **spotify**: Music streaming (search songs, playlists, recommendations)
- **venmo**: Payment transactions (send/request money, view transactions)
- **gmail**: Email management (send, read, search emails)
- **splitwise**: Expense sharing (create groups, split bills, settle debts)
- **simple_note**: Note-taking (create, edit, search notes)
- **todoist**: Task management (create tasks, projects, set deadlines)
- **api_docs**: API documentation access (discover available APIs)
- **supervisor**: Task management (user info, task completion)

---

## Code Execution Model

**IMPORTANT**: AppWorld uses a **Python code execution paradigm**, not simple tool calls.

### Key Characteristics:
1. **Persistent Runtime**: Variables persist across interactions within the same episode
2. **Python Syntax**: Use standard Python code (loops, conditionals, variable assignments)
3. **Code Delimiters**: Wrap your code in `<code>...</code>` tags
4. **Print Outputs**: You MUST `print()` results to see them, otherwise only execution status is returned

### Example Interaction:
<code>
# Get Spotify recommendations
recommendations = apis.spotify.show_recommendations(access_token=spotify_access_token, page_index=0)
print(recommendations)
</code>

---

## API Discovery Tools

Since there are 400+ APIs, you cannot memorize them all. Use these tools to discover available APIs:

### `apis.api_docs.show_api_descriptions(app_name)`

**Description**  
Returns a list of all available APIs for a specific app with brief descriptions.

**Parameters**
- `app_name` (str): Name of the app (e.g., 'spotify', 'gmail', 'amazon')

**Usage Example**
<code>
# Discover all Spotify APIs
spotify_apis = apis.api_docs.show_api_descriptions(app_name='spotify')
print(spotify_apis)
</code>

**Returns**  
A list of API names and their brief descriptions for the specified app.

---

### `apis.api_docs.show_api_doc(app_name, api_name)`

**Description**  
Returns detailed documentation for a specific API, including parameters, return values, and usage examples.

**Parameters**
- `app_name` (str): Name of the app (e.g., 'spotify', 'gmail', 'amazon')
- `api_name` (str): Name of the specific API (e.g., 'show_recommendations', 'send_email')

**Usage Example**
<code>
# Get detailed documentation for a specific API
doc = apis.api_docs.show_api_doc(app_name='spotify', api_name='show_recommendations')
print(doc)
</code>

**Returns**  
Detailed documentation including parameter types, descriptions, and usage examples.

---

## User Context and Credentials

There are many tools to be used to get information and credentials for the user. These tools are under the 'supervisor' 
API: 
['show_active_task', 'complete_task', 'show_profile', 'show_addresses', 'show_payment_cards', 'show_account_passwords']. 
We give below an example for the 'show_account_passwords' tool.

### `apis.supervisor.show_account_passwords()`

**Description**  
Returns password information about the current user.

**Usage Example**
<code>
passwords = apis.supervisor.show_account_passwords()
print(passwords)
# Extract credentials for use in other API calls
spotify_password = passwords['spotify']
</code>


---
## Task Completion

### `apis.supervisor.complete_task(answer=None)`

**Description**  
Signals task completion and, if needed, submits the final answer. This is the ONLY way to complete a task.

**Parameters**
- `answer` (str, optional): The final answer to the task question. Some tasks are evaluated based on environment state
rather than the answer string, or both. If the task does not require an answer, leave this as None. If the task 
requires an answer, provide it as a string. If the task asks for a number, only provide that number as a string, 
without additional information. If the task asks for a name, only provide that name as a string, without additional 
information. If the task asks for a Yes/No question, only output 'Yes' or 'No', without additional information.
In general, strictly avoid unnecessary text in the answer.
Only provide the exact required information.

**Usage Example with answer**
<code>
# Complete the task with an answer
apis.supervisor.complete_task(answer="Albert Einstein")
</code>

**Usage Example without answer**
<code>
# Complete the task without an answer
apis.supervisor.complete_task()
</code>

**Important Notes**
- Task completion triggers evaluation tests
- Evaluation may check the answer string AND/OR the environment state
- For example: "Was the email sent to the right person with the right content?"
- You receive reward=1.0 only if ALL evaluation tests pass and the answer (if needed) is correct


---
## Execution Guidelines
1. **Start with Discovery**: Use `apis.api_docs.show_api_descriptions()` to find relevant APIs for the task
2. **Get Details**: Use `apis.api_docs.show_api_doc()` for specific API documentation
3. **Get Credentials**: Use `apis.supervisor` tools ()'show_active_task', 'complete_task', 'show_profile', 
'show_addresses',  'show_payment_cards', 'show_account_passwords') to obtain necessary information, 
passwords, addresses, etc. about the user.
4. **Authenticate**: All Apps (except for 'supervisor' and 'api_docs') require authentication to use their APIs. 
Use obtained credentials to authenticate with the 'login' API for the target app. 
This will return an access_token for subsequent API calls in that app.
4. **Execute Operations**: Write Python code to accomplish the task using API calls
5. **Print Results**: Always `print()` outputs to see results
6. **Complete Task**: Call `apis.supervisor.complete_task()` when done

### Example trajectory
Question: "Name the artist most recommended to me on Spotify."

Code generated across ReAct steps:

# Step 1: Discover available Spotify APIs
<code>
spotify_apis = apis.api_docs.show_api_descriptions(app_name='spotify')
print(spotify_apis)
</code>

# Step 2: Get detailed documentation for specific API
<code>
doc = apis.api_docs.show_api_doc(app_name='spotify', api_name='show_recommendations')
print(doc)
</code>

# Step 3: Get user credentials (email)
<code>
user = apis.supervisor.show_profile()
print(user)
</code>

# Step 4: Get user credentials (password)
<code>
passwords = apis.supervisor.show_account_passwords()
print(passwords)
</code>

# Step 5: Login and extract the access token
<code>
password = passwords['spotify']
username = user['email']
login_result = apis.spotify.login(username=username, password=password)
print(login_result)
</code>

# Step 6: Find spotify recommendations
<code>
spotify_access_token = login_result['access_token']
recommendations = apis.spotify.show_recommendations(
    access_token=spotify_access_token,
    page_index=0
)
print(recommendations)
</code>

# Step 7: Find most recommended artist
<code>
from collections import Counter
artist_counter = Counter()
for recommendation in recommendations:
    for artist in recommendation["artists"]:
        artist_counter[artist["name"]] += 1
most_recommended_artist = artist_counter.most_common(1)[0][0]
print(most_recommended_artist)
</code>

# Step 8: Complete the task
<code>
apis.supervisor.complete_task(answer=most_recommended_artist)
</code>


---
## Advanced Python Usage
You can use standard Python constructs:

### Loops
<code>
# Process multiple items
for i in range(5):
    result = apis.spotify.add_song_to_playlist(playlist_id=playlist_id, song_id=song_ids[i], access_token=access_token)
    print(f"Track {i}: {result}")
</code>

### Conditionals
<code>
# Conditional logic
if balance > 100:
    apis.venmo.create_transaction(amount=50, receiver_email="friend@venmo", access_token=access_token)
else:
    print("Insufficient balance")
</code>

### Variable Reuse
<code>
# Variables persist across interactions
playlist_id = apis.spotify.create_playlist(name="My Favorites", access_token=access_token)
print(f"Created playlist: {playlist_id}")
</code>

Next interaction:
<code>
# Reuse the playlist_id from previous interaction
apis.spotify.add_song_to_playlist(playlist_id=playlist_id, song_id="abc123", access_token=access_token)
print("Track added to playlist")
</code>


---
## Important Reminders
1. **Always use `<code>...</code>` delimiters** around your Python code
2. **Always `print()` results** to see outputs (otherwise you only see execution status)
3. **Use API discovery tools** to find the right APIs for your task
4. **Get user credentials** before making API calls that require authentication
5. **Complete the task** with `apis.supervisor.complete_task()` when done
6. **Evaluation is comprehensive**: Both your answer and environment state are checked
```
\end{verbatim}
}
\end{tcolorbox}
\captionsetup{hypcap=false}
\captionof{figure}{Executor tool prompt for the AppWorld environment.}
\captionsetup{hypcap=true}
\label{fig:appworld-prompt}
\end{center}

\subsection{Wikipedia Environment}
\label{app:wikienv}

We implement a lightweight interactive environment that enables an agent to consult Wikipedia during multi-step reasoning. 
The design is based on the \emph{Wikienv} used in the ref.~\cite{Yao2022-nl}, with some minor improvements such as improved parsing of tables and Wikipedia's sideboxes (e.g. those containing biographical data)

\subsubsection{Environment Overview}

The environment exposes Wikipedia as a read-only external knowledge source through a small set of tools.  
Each episode proceeds as a sequence of tool invocations and terminates only when the \texttt{finish} action is called.

At any step, the agent observes the textual output of the tool it invoked.  
The environment maintains implicit state in the form of the currently referenced Wikipedia page, enabling incremental lookup without requiring the agent to repeatedly re-issue full queries.

\subsubsection{Available Tools}

The environment supports exactly three tools:
\begin{itemize}
    \item \texttt{search[entity]}: retrieves introductory information about a Wikipedia entity.
    \item \texttt{lookup[string]}: retrieves the next sentence from the current page that matches a query string.
    \item \texttt{finish[answer]}: terminates the episode and returns the final answer.
\end{itemize}

Only one tool may be called per turn, and tools must be invoked explicitly by name.

\subsubsection{Tool Prompt}

Listing~\ref{fig:wikienv-prompt} shows the full tool prompt. This prompt defines the available tools, their behavior, and usage constraints; when the an agent operates in the Wikienv (i.e. for the HotpotQA and FanOutQA benchmarks) this prompt is inlined into the "Available Tools" section of all the agent prompts we show in Appendix~\ref{app:prompts}.

\begin{center}
\begin{tcolorbox}[
  width=\linewidth,
  breakable,
  colback=blue!5,
  colframe=blue!60!black,
  boxrule=0.6pt,
  left=3pt,
  right=3pt,
  top=3pt,
  bottom=3pt,
  title={Tool description for the Wikienv environment},
  fonttitle=\bfseries,
  sharp corners
]
{\tiny
\begin{verbatim}
### `search[entity]`

**Description**  
Retrieves introductory information about a Wikipedia entity.

**Behavior**  
- Returns the **first five sentences** from the Wikipedia page corresponding to `entity`.

**Typical Use Cases**
- Obtain a high-level overview of a topic.
- Identify key facts or context before deeper lookup.

**Example usage**
Let's assume that the query is "Did Richard Feynman win a Nobel Prize?"
You would then search for his Wikipedia page like so

Tool call: search[Richard Feynman]
Output: Richard Phillips Feynman (May 11, 1918 – February 15, 1988)[1]
was an American theoretical physicist. He shared the 1965 Nobel Prize
in Physics with Julian Schwinger and Shin'ichirō Tomonaga \"for their
fundamental work in quantum electrodynamics (QED), with deep-ploughing
consequences for the physics of elementary particles\".[2]. He is also
known for his work in the path integral formulation of quantum mechanics,
the theory of the physics of the superfluidity of supercooled liquid helium,
and the parton model. Feynman developed a pictorial representation scheme
for the mathematical expressions describing the behavior of subatomic particles,
which later became known as Feynman diagrams and is widely used..He assisted
in the development of the atomic bomb during World War II and became known to the wider 
public in the 1980s as a member of the Rogers Commission, the panel 
that investigated the Space Shuttle Challenger disaster.

---

### `lookup[string]`

**Description**  
Retrieves additional information from the currently referenced Wikipedia page.

**Behavior**  
- Returns the **next sentence** in the page that contains the provided `string`.

**Typical Use Cases**
- Find specific facts or details mentioned later in the article.
- Incrementally gather information after an initial `search`.

**Example usage**
You would then lookup "Nobel Prize" in his bio, like so:
Tool call: lookup[Nobel Prize]
Output: (Result 1 / 11) Richard Phillips Feynman (May 11, 1918 – February 15, 1988)[1]
was an American theoretical physicist. He shared the 1965 Nobel Prize in Physics with
Julian Schwinger and Shin'ichirō Tomonaga \"for their fundamental work in quantum 
electrodynamics (QED), with deep-ploughing consequences for the physics of elementary
particles\".[2]

Tool call: lookup[Nobel Prize]
Output: (Result 2 / 11) Feynman's parents were both from Jewish families,[5] and his family 
went to the synagogue every Friday.[18] However, by his youth, Feynman described himself as
an \"avowed atheist\".[19][20] Many years later, in a letter to Tina Levitan, declining a
request for information for her book on Jewish Nobel Prize winners, he stated, \"To select,
for approbation the peculiar elements that come from some supposedly Jewish heredity is to
open the door to all kinds of nonsense on racial theory\", adding, \"at thirteen I was not
only converted to other religious views, but I also stopped believing that the Jewish people
are in any way 'the chosen people'\".[21]

---

### `finish[answer]`

**Description**  
Terminates the episode and returns the final answer to the user.

**Behavior**  
- Outputs `answer` as the final response.
- Ends the interaction; no further tool calls are allowed.

**Typical Use Cases**
- Deliver the synthesized answer once all required information has been gathered.

**Example usage**
You would then provide an answer in the affirmative (because Richard Feynman
*did* win a Nobel Prize in Physics), like so:
Tool call: finish[yes]
\end{verbatim}
}
\end{tcolorbox}
\captionsetup{hypcap=false}
\captionof{figure}{Executor tool prompt for the Wikipedia environment.}
\captionsetup{hypcap=true}
\label{fig:wikienv-prompt}
\end{center}

\section{Qualitative Examples}
\subsection{PEVR replans vs EVA advices in AppWorld}
\label{app:pevr_eva_appworld}
We report in Figure~\ref{fig:pevr_replan_appworld} and Figure~\ref{fig:eva_advice_appworld} respectively the replan from PEVR and the advice from EVA for a task in AppWorld. We observe a stark comparison in how actionable and detailed the replan from PEVR is in comparison with the generalist advice from EVA.

\begin{center}
\begin{tcolorbox}[
  width=\linewidth,
  breakable,
  colback=green!5,
  colframe=green!55!black,
  boxrule=0.6pt,
  left=3pt,
  right=3pt,
  top=3pt,
  bottom=3pt,
  title={Qualitative Example of PEVR replan in AppWorld},
  fonttitle=\bfseries,
  sharp corners
]
{\scriptsize
\begin{verbatim}
INTERVENE
<REPLAN>
1. **Discover Amazon APIs**  
   - Use `apis.api_docs.show_api_descriptions(app_name='amazon')` to retrieve a list of available APIs 
   for interacting with the Amazon app.  
   - Expected output: A list of API names and their brief descriptions. This will help identify the API
   to fetch the cart details.

2. **Get Detailed Documentation for Cart API**  
   - Use `apis.api_docs.show_api_doc(app_name='amazon', api_name='[CART_API_NAME]')` (replace 
   `[CART_API_NAME]` with the identified API from step 1) to understand the inputs and outputs required
   for retrieving cart details.  
   - Expected output: Detailed documentation for the cart API, including parameter requirements and
   example responses.

3. **Discover User Credentials for Amazon**  
   - Use `apis.supervisor.show_account_passwords()` to retrieve the user's stored Amazon credentials.  
   - Expected output: A dictionary containing account credentials, including the Amazon account 
   username and password.

4. **Log in to Amazon**  
   - Use the retrieved credentials to log in to the Amazon app using the `login` API.  
   - Expected input: Username and password for the Amazon account.  
   - Expected output: An `access_token` to authenticate subsequent API calls to Amazon.

5. **Retrieve Cart Details**  
   - Use the identified cart API from step 2 with the `access_token` obtained in step 4 to fetch
   the cart details.  
   - Expected input: The user's `access_token`.  
   - Expected output: A list of items in the cart, including their names and prices.

6. **Calculate Total Cart Cost**  
   - Parse the retrieved cart details to calculate the total cost of all items in the cart, 
   excluding tax and delivery fees.  
   - Expected output: The total cost as a numeric value.

7. **Discover Venmo APIs**  
   - Use `apis.api_docs.show_api_descriptions(app_name='venmo')` to retrieve a list of available
   APIs for interacting with the Venmo app.  
   - Expected output: A list of API names and their brief descriptions. This will help identify
   the API for requesting money.

8. **Get Detailed Documentation for Money Request API**  
   - Use `apis.api_docs.show_api_doc(app_name='venmo', api_name='[REQUEST_API_NAME]')` (replace
   `[REQUEST_API_NAME]` with the identified API from step 7) to understand the inputs and outputs 
   required for requesting money.  
   - Expected output: Detailed documentation for the money request API, including parameter
   requirements and example responses.

9. **Discover User Credentials for Venmo**  
   - Use `apis.supervisor.show_account_passwords()` to retrieve the user's stored Venmo credentials.  
   - Expected output: A dictionary containing account credentials, including the Venmo account 
   username and password.

10. **Log in to Venmo**  
    - Use the retrieved credentials to log in to the Venmo app using the `login` API.  
    - Expected input: Username and password for the Venmo account.  
    - Expected output: An `access_token` to authenticate subsequent API calls to Venmo.

11. **Request Money from Adam**  
    - Use the identified money request API from step 8 with the `access_token` obtained in step
    10 to request the calculated total cart cost from Adam.  
    - Expected input: Adam's Venmo username or email, the calculated total cost (from step 6), 
    and a note indicating the reason for the request (e.g., "Reimbursement for Amazon cart items").  
    - Expected output: Confirmation that the money request has been sent successfully.

12. **Complete the Task**  
    - Use `apis.supervisor.complete_task()` to signal task completion. No specific answer needs 
    to be provided, as the task is evaluated based on the successful execution of the steps.  
    - Expected output: Task completion confirmation.
</REPLAN>
\end{verbatim}
}
\end{tcolorbox}
\captionsetup{hypcap=false}
\captionof{figure}{Qualitative Example of PEVR replan in AppWorld. The replan is detailed and actionable, discussing expected inputs and outputs for each step of the plan.}
\captionsetup{hypcap=true}
\label{fig:pevr_replan_appworld}
\end{center}

\begin{center}
\begin{tcolorbox}[
  width=\linewidth,
  colback=green!5,
  colframe=green!55!black,
  boxrule=0.6pt,
  left=3pt,
  right=3pt,
  top=3pt,
  bottom=3pt,
  title={Qualitative Example of EVA advice in AppWorld},
  fonttitle=\bfseries,
  sharp corners,
  breakable
]
{\scriptsize
\begin{Verbatim}[breaklines=true, breakanywhere=true, breakautoindent=true, breaksymbolleft=\tiny\ensuremath{\hookrightarrow}\,]
assistant:
INTERVENE
<SUMMARY>
The Executor has repeatedly attempted to use nonexistent API names for accessing the Amazon cart 
subtotal, such as 'get_cart', 'get_cart_items', 'get_cart_summary', and 'get_cart_total'. Each 
attempt resulted in a failure due to a 422 error, indicating the API name provided was invalid. 
No progress has been made toward retrieving the cart subtotal or requesting payment through Venmo.
</SUMMARY>
<ADVICE>
1. Use `apis.api_docs.show_api_descriptions('amazon')` to list the available APIs for Amazon.
2. Identify the correct API name for accessing the cart subtotal or items from the list returned.
3. Authenticate with Amazon using credentials from supervisor tools, retrieve the access token, 
and call the identified API to get the cart subtotal.
4. Authenticate with Venmo using credentials from supervisor tools, retrieve the access token, 
and use the `request_money` API to send the payment request to Adam.
Corrections:
- Avoid guessing API names; rely on `show_api_descriptions` to confirm available APIs.
- Ensure the correct access tokens are used for each platform (Amazon vs. Venmo).
- Provide consistent and valid code blocks to avoid syntax errors or incomplete commands.
</ADVICE>
\end{Verbatim}
}
\end{tcolorbox}
\captionsetup{hypcap=false}
\captionof{figure}{Qualitative Example of EVA advice in AppWorld. The advice is general and synthetic. While it provides some useful advice on pitfalls to avoid, it fails to give a concrete path with next steps for the executor.}
\captionsetup{hypcap=true}
\label{fig:eva_advice_appworld}
\end{center}

\section{Hardware}
\label{app:hardware}
As mentioned in Section~\ref{subsec:models}, all experiments were performed using a cluster of NVIDIA A100 GPUs relying on a single GPU.
For cloud models, we used the standard subscription service provided by Azure OpenAI for GPT-4o.
Shortest individual experiments completed in 1-2 hours, while longer experiments on AppWorld test set with Qwen 3 32B models took up to 1-2 days.

\section{Societal Impact}
\label{app:impact}
Our research does not introduce novel methods, nor considers specific applications which could have potential positive or negative societal impact. We believe the only societal impact from our study could be to positively influence the experience and monetary costs of the consumers of LLM agents, as we show that Hybrid Multi-Agent systems enable a better trade-off between cloud subscription costs and power consumption of edge devices.

\section{LLM Usage}
\label{app:llm}
In the scope of this submission, LLMs have been used for the initial literature review, and to help write individual parts of the Python codebase. We highlight that the code written by LLMs was peer-reviewed by multiple expert human engineers before being merged and used for experimentation.

\end{document}